\documentclass[a4paper,11pt]{article}
\pdfoutput=1
\usepackage{jheppub}

\usepackage{epsf}
\usepackage{epsfig}
\usepackage{subfigure}
\usepackage{mathtools}
\usepackage{hhline}
\usepackage{float}
\usepackage{multirow}
\usepackage{nicefrac}
\usepackage{epstopdf}
\usepackage{slashed}
\usepackage{xcolor}
\usepackage{url}
\usepackage{umoline}

\newcommand{\dis}[1]{\begin{equation}\begin{split}#1\end{split}\end{equation}}

\newcommand{\be}{\begin{eqnarray}}
\newcommand{\ee}{\end{eqnarray}}

\newcommand{\ba}{\begin{array}}
\newcommand{\ea}{\end{array}}
\newcommand{\bee}{\begin{equation}\ba{c}}
\newcommand{\eee}{\ea\end{equation}}

\newcommand{\bi}{\begin{itemize}}
\newcommand{\ei}{\end{itemize}}

\toccontinuoustrue

\title{Two Higgs doublet model with vectorlike leptons and contributions to $pp\to WW$ and $H\to WW$}
\author{Radovan Derm\'i\v{s}ek$^{1,2}$,}
\author{Enrico Lunghi$^1$}
\author{and Seodong Shin$^1$}
\affiliation{
$^1$Physics Department, Indiana University, Bloomington, IN 47405, USA \\
$^2$Department of Physics and Astronomy and Center for Theoretical Physics, Seoul National University, Seoul 151-747, Korea \\
}
\emailAdd{dermisek@indiana.edu} 
\emailAdd{elunghi@indiana.edu} 
\emailAdd{shinseod@indiana.edu}

\abstract{We study a two Higgs doublet model  extended by  vectorlike leptons mixing with one family of standard model leptons. Generated  flavor violating couplings between heavy and light leptons can dramatically alter the decay patterns of heavier Higgs bosons. We focus on $pp \to H \to \nu_4 \nu_\mu \to W \mu \nu_\mu$, where $\nu_4$ is a new neutral lepton,  and study possible effects of this process on the measurements of $pp \to WW$ and $H \to WW$  since it leads to the same final states.   We discuss predictions for  contributions to $pp \to WW$ and $H \to WW$  and their correlations from the region of the parameter space that satisfies all available constraints including precision electroweak observables and from pair production of vectorlike leptons.  Large contributions, close to current limits, favor small $\tan \beta $ region of the parameter space. We find that, as a result of adopted cuts in experimental analyses, the contribution to  $pp \to WW$  can be an order of magnitude larger than the contribution to $H \to WW$. Thus, future precise measurements of $pp\to WW$ will further constrain the parameters of the model. In addition, we also consider possible contributions to $pp\to WW$ from the heavy Higgs decays into a new charged lepton $e_4$ ($H\to e_4 \mu \to W \mu \nu_\mu$), exotic SM Higgs decays, and pair production of vectorlike leptons.

\newpage
 
}
\preprint{
\begin{minipage}{3cm}
\small
\flushright
IU-HET-602
\end{minipage}} 

\begin{document}

\maketitle 

\section{Introduction}
\label{sec:introduction}
Among simple extensions of the standard model (SM) are those with extended Higgs sector and extra vectorlike leptons near the electroweak (EW) scale. Since masses of vectorlike leptons are not related to  their Yukawa couplings, in the absence of mixing with SM leptons, they are not strongly constrained by experiments. However, even small Yukawa couplings between  SM leptons and vectorlike leptons can significantly affect a variety of processes and can dramatically alter the decay patterns of heavier Higgs bosons.
  
We consider an extension of the two Higgs doublet model type-II by vectorlike pairs of new leptons: SU(2) doublets $L_{L,R}$,  SU(2) singlets $E_{L,R}$ and  SM singlets $N_{L,R}$, where  $L_L$ and $E_R$ have the same hypercharges as SM leptons. We further assume that the new leptons mix only with one family of SM  leptons and we consider the mixing with the second family as an example. The mixing of new vectorlike leptons with leptons in the SM  generate  flavor violating couplings of  $W$, $Z$ and Higgs bosons between heavy and light leptons. These couplings can result in new decay modes for heavy CP even (or CP odd) Higgs boson: $H \to \nu_4 \nu_\mu$ and $H \to e_4 \mu$, where $e_4$ and $\nu_4$ are the  lightest new charged and neutral leptons. These decay modes, when kinematically open, can be very important, especially when the mass of  the heavy Higgs boson is below the $t \bar t$ threshold, about 350 GeV, and the light Higgs boson ($h$) is SM-like so that $H \to ZZ,\; WW$ are suppressed or not present. In this case, flavor violating decays $H \to \nu_4 \nu_\mu$ or $H \to e_4 \mu$ compete only with $H\to b \bar b$ (for sufficiently heavy $H$ also with $H \to hh$) and can be large or even dominant. 
Subsequent decay modes: $e_4 \to W  \nu_\mu$, $e_4 \to Z \mu$,  $e_4 \to h \mu$ and $\nu_4 \to W \mu$, $\nu_4 \to Z \nu_\mu$, $\nu_4 \to h \nu_\mu$ lead to many possible final states.

In this paper, we focus on $pp \to H \to \nu_4 \nu_\mu \to W \mu \nu_\mu$ and study possible effects of this process on the measurements of $pp \to WW$ and $H \to WW$  since it leads to the same final states. This process  was previously studied by us in a model independent way~\cite{Dermisek:2015vra} in the connection with the ATLAS excess in  $pp \to WW$~\cite{atlasww}. The results were presented in terms of the Higgs mass, the mass of $\nu_4$ and the product of branching ratios ${\rm BR}(H \to \nu_4 \nu_\mu) \; {\rm BR}(\nu_4 \to W \mu)$. Here we study the process in detail in the two Higgs doublet model type-II which is perhaps the simplest realization of the scenario. We discuss predictions for  contributions to $pp \to WW$ and $H \to WW$  and their correlations from the region of the parameter space that satisfies all available constraints including precision electroweak observables~\cite{Agashe:2014kda} and constraints from pair production of vectorlike leptons~\cite{Dermisek:2014qca}.  Large contributions, close to current limits, favor small $\tan \beta $ region of the parameter space. We find that, as a result of adopted cuts in experimental analyses, the contribution to  $pp \to WW$  can be an order of magnitude larger than the contribution to $H \to WW$. In addition, we also consider  possible contributions to $pp\to WW$ from $H\to e_4 \mu \to W \mu \nu_\mu$, from similar processes involving  SM-like Higgs boson and from pair production of vectorlike leptons.

Vectorlike leptons near the electroweak scale provide a very rich phenomenology and were studied in a variety of contexts. Most of the previous studies would apply also to the two Higgs doublet model we consider here since we assume type-II couplings of Higgs doublets to fermions relevant for supersymmetric extensions and we also consider the limit when the light Higgs is SM-like which is relevant for SM extensions by vectorlike fermions. For example, analogous processes involving  SM-like Higgs boson decaying into $2\ell 2\nu$ or $4 \ell$ through a new lepton were previously studied in ref.~\cite{Dermisek:2014cia} and the  $4\ell$ case also in  ref.~\cite{Falkowski:2014ffa}. Possible explanation of the muon g-2 anomaly with vectorlike leptons was studied in \cite{Kannike:2011ng, Dermisek:2013gta}. Further extensions with vectorlike quarks and possibly $Z'$ are straightforward and offer possibilities to explain anomalies in Z-pole observables~\cite{Choudhury:2001hs, Dermisek:2011xu, Dermisek:2012qx, Batell:2012ca}. In addition, extensions with complete vectorlike families were considered that provide an understanding of  values of gauge couplings from IR fixed point behavior and threshold effects of vectorlike fermions, as in insensitive unification~\cite{Dermisek:2012as, Dermisek:2012ke}.  Many studies were also done in supersymmetric framework, see for example~Refs.~\cite{Babu:1996zv, BasteroGil:1999dx, Kolda:1996ea, Barr:2012ma, Martin:2009bg, Bae:2012ir}, and  in various  frameworks the constraints from precision  electroweak data have been analyzed~\cite{Lavoura:1992np,Dawson:2012di,Joglekar:2012vc,Kearney:2012zi,Garg:2013rba,Altmannshofer:2013zba,Dorsner:2014wva}. Further discussion and more references can be found in a recent review~\cite{Ellis:2014dza}.

This paper is organized as follows. In section~\ref{sec:model} we present two Higgs doublet model type-II with vectorlike leptons mixing with one family of the SM leptons and derive formulas for couplings of $Z$, $W$ and Higgs bosons to leptons. In section~\ref{sec:BRs} we discuss branching ratios of the heavy Higgs boson $H$ and neutral lepton $\nu_4$ and find approximate expressions assuming small mixing between heavy and light leptons. The scan over the parameter space of the model is described in section~\ref{sec:constraints} together with constraints imposed from precision electroweak data and direct searches for new leptons. The approximate formulas derived in previous sections are useful for understanding the main results which are presented in section~\ref{sec:results}. For completeness, we also discuss possible contributions to $pp\to WW$ from $H\to e_4 \mu \to W \mu \nu_\mu$ in section~\ref{sec:e4}, from the SM-like Higgs boson in section~\ref{sec:SMh} and from pair production of vectorlike leptons in section~\ref{sec:DY}. We summarize and present  concluding remarks in section~\ref{sec:conclusions}.

\section{Model}
\label{sec:model}
We consider an extension of a two Higgs doublet model by vectorlike pairs of new leptons: SU(2) doublets $L_{L,R}$,  SU(2) singlets $E_{L,R}$ and  SM singlets $N_{L,R}$. The quantum numbers of  new particles are summarized in Table \ref{table:fieldcontents}.  The  $L_L$ and $E_R$ have the same quantum numbers as the muon doublet $\mu_L$ (we use the same label for the charged component as for the whole doublet) and the right-handed muon $\mu_R$ respectively.  We further assume that the new leptons mix only with one family of SM  leptons and we consider the mixing with the second family as an example. This can be achieved by requiring that the individual lepton number is an approximate symmetry (violated only by light neutrino masses). The results for mixing with the first or the third family could be obtained in the same way. The  mixing of new leptons  with more than one SM family simultaneously is strongly constrained by various lepton flavor violating processes and we will not pursue this direction here. Finally, we assume that leptons couple to the two Higgs doublets as in the type-II model, namely the down sector couples to $H_d$ and the up sector couples to $H_u$. This can be achieved by the $Z_2$ symmetry specified in Table \ref{table:fieldcontents}. The generalization to the whole vectorlike family of new leptons, including the quark  sector, would be straightforward. 
\begin{table}[htp]
    \caption{Quantum numbers of  standard model leptons, extra vectorlike leptons and the two Higgs doublets. The electric charge is given by $Q = T_3 +Y$, where $T_3$ is the weak isospin, which is +1/2 for the first component of a doublet and -1/2 for the second component. }
\begin{center}
\begin{tabular}{c c c c c c c c}
\hline
\hline
 & ~~$\mu_L$ & ~~$\mu_R$ & ~~$L_{L,R}$ & ~~$E_{L,R}$ & ~~$N_{L,R}$ & ~~$H_d$ & ~~ $H_u$\\
\hline
SU(2)$_{\rm L}$ & ~~\bf 2 & ~~\bf 1 & ~~\bf 2 & ~~\bf 1 & ~~\bf 1 & ~~\bf 2 & ~~\bf 2 \\
U(1)$_{\rm Y}$ & ~~-$\frac12$ & ~~-1 & ~~-$\frac12$ & ~~-1 & ~~0 & ~~$\frac12$ & ~~-$\frac12$ \\
Z$_2$ & ~~+ & ~~-- & ~~+ & ~~-- & ~~+ & ~~-- & ~~+ \\
\hline
\hline
\end{tabular}
\end{center}
\label{table:fieldcontents}
\end{table}

With these assumptions, the most general renormalizable Lagrangian containing Yukawa and mass terms for the second generation of SM leptons and new vectorlike leptons is given by:
\dis{
{\cal L} \supset \;  & - y_{\mu} \bar \mu_{L}  \mu_{R} H_d - \lambda_E \bar \mu_{L}  E_{R} H_d  -  \lambda_L \bar L_{L} \mu_{R} H_d -  \lambda \bar L_{L}  E_{R} H_d - \bar \lambda H_d^\dagger \bar E_{L}  L_{R}  \\
& - \kappa_N  \bar \mu_L  N_R H_u - \kappa  \bar{L}_L  N_R H_u - \bar \kappa H_u^\dagger \bar{N}_L L_R  \\
&  - M_L \bar L_L L_R - M_E \bar E_L E_R - M_N \bar N_L N_R + {\rm h.c.}~,
\label{eq:lagrangian}
}
where the first term is the usual SM Yukawa coupling of the muon, followed by  Yukawa couplings to $H_d$ (denoted by various $\lambda$s) that will result in  masses and couplings of the charged leptons,  Yukawa couplings to $H_u$ (denoted by various $\kappa$s) that will result in  masses and couplings of the neutral leptons, and finally mass terms for vectorlike  leptons. The components of doublets are labeled as follows:
\dis{
\mu_L  = \left( 
\begin{array}{c}
\nu_\mu \\
\mu^-_L
\end{array}
\right),~
L_{L,R}  = \left( 
\begin{array}{c}
L_{L,R}^0 \\
L_{L,R}^-
\end{array}
\right),~
H_d  = \left( 
\begin{array}{c}
H_d^+ \\
H_d^0 
\end{array}
\right),~
H_u  = \left( 
\begin{array}{c}
H_u^0\\
H_u^- 
\end{array}
\right),
}
where the neutral Higgs components develop the vacuum expectation values $\left< H_u^0 \right> = v_u$ and $\left< H_d^0 \right> = v_d$. We assume that both are real and positive as in the $CP$ conserving two Higgs doublet model with $\sqrt{v_u^2 + v_d^2} = v = 174$ GeV and we define $\tan \beta \equiv v_u / v_d$. 

After spontaneous symmetry breaking the resulting mass matrices in the charged and neutral sectors  can be diagonalized and we label the two new charged and neutral mass eigenstates by $e_4$ and $e_5$ and $\nu_4$ and $\nu_5$ respectively. Couplings off all involved particles to the $Z$, $W$ and Higgs bosons are in general modified  because  SU(2) singlets mix with SU(2) doublets. The flavor conserving couplings receive corrections and flavor violating couplings between the muon (or muon neutrino)  and heavy leptons are generated. The couplings resulting from the mixing in the charged sector were discussed in detail in ref.~\cite{Dermisek:2013gta} in the connection with the muon g-2 anomaly. Here we will focus on couplings resulting from the mixing in the neutral sector. These are also more relevant for the discussion of the contribution of the Higgs boson decays to $pp \to WW$.

The mass matrix in the neutral lepton sector is given by: 
\dis{
\left( 
\begin{array}{ccc}
\bar \nu_\mu & \bar{L}_L^0 & \bar N_L
\end{array}
\right)
M_\nu 
\left( 
\begin{array}{c}
\nu_R = 0 \\
L_R^0 \\
N_R
\end{array}
\right) = 
\left( 
\begin{array}{ccc}
\bar \nu_\mu & \bar{L}_L^0 & \bar N_L
\end{array}
\right)
\left( 
\begin{array}{ccc}
0 & 0 & \kappa_N v_u \\
0 & M_L & \kappa v_u \\
0 & \bar \kappa v_u & M_N \\
\end{array}
\right)
\left(
\begin{array}{c}
\nu_R = 0 \\
L_R^0 \\
N_R
\end{array}
\right)~, 
\label{eq:mm}
}
where we inserted $\nu_R = 0$ for the  right-handed neutrino which is absent in our framework in order to keep  the mass matrix $3\times3$ in complete analogy with the charged sector. For the discussion of couplings it is convenient to define vectors $\nu_{La} \equiv ( \nu_\mu, {L}_L^0, N_L)^T$ and $\nu_{Ra} \equiv ( \nu_R = 0, {L}_R^0, N_R)^T$. The mass matrix $M_\nu$ can be diagonalized by a biunitary transformation 
\dis{
V_L^\dagger M_\nu V_R = 
\left(
\begin{array}{c c c}
 0& 0 & 0 \\
0 & m_{\nu_4} & 0 \\
0 & 0 & m_{\nu_5}
\end{array}
\right)~,
\label{eq:VLMVR}
}
resulting in masses for  $\nu_4$ and $\nu_5$ leaving  the muon neutrino massless.  The light neutrino masses can be generated by a variety of ways. Once they are generated, the mixing of light neutrinos with vectorlike leptons results in  corrections to both the masses and mixing angles controlled by Yukawa couplings in eq.~(\ref{eq:lagrangian}). 

For better understanding of corrections to gauge and Yukawa couplings discussed later, approximate analytic formulas for diagonalization matrices are useful. These can be obtained in analogy with those in the charged lepton sector given in ref.~\cite{Dermisek:2013gta}. In the limit
\dis{
\kappa_N v_u, \kappa v_u, \bar \kappa v_u \ll M_L, M_N
\label{eq:templimit}
}
with $M_L$ and $M_N$ not close to each other, we find
\dis{
V_L &= 
\left(
\begin{array}{c c c}
1  - \frac{\kappa_N^2 v_u^2}{2M_N^2} & -\frac{\kappa_N v_u^2}{M_L} \frac{\kappa M_L + \bar \kappa M_N}{M_N^2 - M_L^2} &  \frac{\kappa_N v_u}{M_N} \\
\frac{\kappa_N \bar \kappa v_u^2}{M_L M_N} & ~~1- \frac{(M_L \bar \kappa + M_N \kappa)^2 v_u^2}{2(M_N^2 - M_L^2)^2} & \frac{(M_L \bar \kappa + M_N \kappa) v_u}{M_N^2 - M_L^2} \\
-\frac{\kappa_N v_u}{M_N} & - \frac{(M_L \bar \kappa + M_N \kappa) v_u}{M_N^2 - M_L^2} &  ~~1- \frac{{\kappa_N}^2 v_u^2}{2M_N^2} - \frac{(M_L \bar \kappa + M_N \kappa)^2 v_u^2}{2(M_N^2 - M_L^2)^2} 
\end{array}
\right)~ 
\label{eq:VL}
}
and
\dis{
V_R &=
\left(
\begin{array}{ccc}
1 & 0 & 0 \\
0 & ~~1- \frac{(M_L \kappa + M_N \bar \kappa)^2 v_u^2}{2(M_N^2  - M_L^2)^2} & \frac{(M_L \kappa + M_N \bar \kappa) v_u}{M_N^2  - M_L^2} \\
0 & -\frac{(M_L \kappa + M_N \bar \kappa) v_u}{M_N^2  - M_L^2} & ~~1 -\frac{(M_L \kappa + M_N \bar \kappa)^2 v_u^2}{2(M_N^2  - M_L^2)^2} \\
\end{array}
\right)
}
up to corrections of $\mathcal{O}(\epsilon^3)$ where $\epsilon = (\kappa_N, \kappa, \bar \kappa) v_u /(M_L,M_N)$. The mass eigenvalues are $0, M_L + \mathcal{O}(\epsilon^2), M_N + \mathcal{O}(\epsilon^2)$. However, in our numerical analysis we do not use any approximations.

\subsection{Couplings of the Z and W bosons}
Couplings of the muon and new heavy leptons  to the $Z$ and $W$ bosons  are modified from their SM values because SU(2) singlets mix with SU(2) doublets. These couplings can be written in terms of $V_L$ and $V_R$, defined in eq.~(\ref{eq:VLMVR}), and of the analogue matrices $U_L$ and $U_R$ that are related to the charged lepton sector and that were discussed in detail in ref.~\cite{Dermisek:2013gta} (with the replacement $v \to v_d$ due to the two Higgs doublet model). The couplings of the $Z$ boson to charged leptons can be found in ref.~\cite{Dermisek:2013gta} and those to neutral leptons follow from  the kinetic terms:
\begin{eqnarray}
{\cal L}_{kin} \supset   &&\bar \nu_{La} i \slashed D_a \nu_{La} +  \bar \nu_{Ra} i \slashed D_a \nu_{Ra}  \nonumber \\
&& = \bar {\hat \nu}_{L a} (V^\dagger_L)_{a c} i \slashed D_c (V_L)_{ cb} \hat \nu_{L b} +  \bar {\hat \nu}_{R a} (V^\dagger_R)_{a c} i \slashed D_c (V_R)_{c b } \hat \nu_{R b},
\label{eq:kinN}
\end{eqnarray}
where the vectors of mass eigenstates are $\hat \nu_{La} \equiv (\hat \nu_\mu,\hat \nu_{L4}, \hat \nu_{L5})^T$ and similarly for $\hat \nu_{Ra} \equiv (\hat \nu_R = 0 ,\hat \nu_{R4}, \hat \nu_{R5})^T$. We label the components of vectors and  diagonalization matrices by 2, 4 and 5 because they correspond to 2nd, 4th and 5th mass eigenstate. The covariant derivative is given by:
\begin{eqnarray}
D_{\mu a} = \partial _\mu 
- i \frac{g}{\cos \theta_W} T^3_a  Z_\mu~, 
\end{eqnarray}
where the weak isospin $T^3_a$ is +1/2 for neutral components of SU(2) doublets and 0 for singlets. Defining couplings of the $Z$ boson to leptons $f_a$ and $f_b$ as
\dis{
\mathcal{L} \supset \left( \bar f_{La} \gamma^\mu g_L^{Z f_a f_b} f_{Lb} + \bar f_{Ra} \gamma^\mu g_R^{Z f_a f_b} f_{Rb} \right) Z_\mu~,
}
we find:
\begin{eqnarray}
g_L^{Z \nu_a \nu_b} &=& \frac{g}{2 \cos \theta_W} \left[ (V_L^\dagger)_{a2} (V_L)_{2b} + (V_L^\dagger)_{a4} (V_L)_{4b} \right]~, \label{eq:gZL}\\
g_R^{Z \nu_a \nu_b} &=& \frac{g}{2 \cos \theta_W} (V_R^\dagger)_{a4} (V_R)_{4b}~. \label{eq:gZR}
\end{eqnarray}
The usual SM couplings of  left-handed neutrinos,
\begin{eqnarray}
(g^{Z\nu_a \nu_b}_L)_{SM} &=& \frac{g}{2\cos \theta_W}   \delta^{ab}
\label{eq:gLSM}
\end{eqnarray}
are modified by
 \begin{eqnarray}
\delta g^{Z\nu_a \nu_b}_L &=& - \frac{g}{2 \cos \theta_W}  (V^\dagger_L)_{a5} (V_L)_{5b}.
\label{eq:delgL}
\end{eqnarray}

The couplings of the  $W$ boson originate from  the kinetic terms:
\begin{eqnarray}
{\cal L}_{kin} \supset \; &&   \frac{g}{\sqrt{2}}  \left(  \bar \nu_{\mu} \gamma^\mu \mu_{L} + \bar L_{L}^0 \gamma^\mu L_{L}^- +   \bar L_{R}^0 \gamma^\mu L_{R}^-  \right)W^+_\mu + h.c.  \nonumber \\
&& =  \frac{g}{\sqrt{2}}  \left(   \bar {\hat \nu}_{L a} (V^\dagger_L)_{a 2} \gamma^\mu (U_L)_{2b} \hat e_{L b} +\bar {\hat \nu}_{L a} (V^\dagger_L)_{a 4}  \gamma^\mu (U_L)_{4b} \hat e_{L b} \right. \nonumber \\
&& \quad \quad\quad \left. +  \; \bar {\hat \nu}_{R a} (V^\dagger_R)_{a 4}  \gamma^\mu (U_R)_{4b} \hat e_{R b}  \right)W^+_\mu + h.c.  ~,
\label{eq:kinW}
\end{eqnarray}
where $\hat e_{La}$, ($\hat e_{Rb}$) are the charged left-handed (right-handed) mass eigenstate and $U_{L}$, $U_R$ are the corresponding diagonalization matrices~\cite{Dermisek:2014cia}. 
Defining couplings of the W boson  to neutrinos $\hat \nu_{a}$ and charged leptons $\hat e_b$ as
\begin{equation}
{\cal L} \supset      \left( \bar {\hat \nu}_{La} \gamma^\mu g^{W\nu_a e_b}_L  \hat e_{Lb} + \bar {\hat \nu}_{Ra} \gamma^\mu g^{W\nu_a e_b}_R  \hat e_{Rb} \right) W^+_\mu + h.c.  ,
\end{equation}
we find:
\begin{eqnarray}
g_L^{W \nu_a e_b} &=& \frac{g}{\sqrt{2}} \left[(V_L^\dagger)_{a2} (U_L)_{2b} + (V_L^\dagger)_{a4} (U_L)_{4b} \right]~,  \label{eq:gWL}\\
g_R^{W \nu_a e_b} &=& \frac{g}{\sqrt{2}} (V_R^\dagger)_{a4} (U_R)_{4b}~. \label{eq:gWR}
\end{eqnarray}

\subsection{Couplings of the Higgs bosons}
As a consequence of explicit mass terms for vectorlike leptons, the usual relations between the mass of a particle and its coupling to  Higgs bosons do not apply. The couplings of neutral  Higgs bosons to neutral leptons can be obtained from the following Yukawa terms in  the Lagrangian (\ref{eq:lagrangian}):
\begin{eqnarray}
{\cal L}_{Y} \supset  && -  \bar \nu_{La}  \, Y_{\nu ab} \, \nu_{Rb} \,  H_u^0   \; + \; h.c. \nonumber \\ && = \; - \bar {\hat \nu}_{L a} (V^\dagger_L)_{a c}  \, Y_{\nu cd} \,  (V_R)_{ db} \, \hat \nu_{R b}  \, H_u^0   \; + \; h.c.,
\end{eqnarray}
where 
\begin{eqnarray}
Y_\nu = 
\begin{pmatrix}
 0 & 0 &  \kappa_N \\
0  &0 &  \kappa \\
 0 & \bar \kappa  & 0
\end{pmatrix}.
\end{eqnarray}
We assume a CP conserving two Higgs doublet model in the limit with the light Higgs $h$ being fully standard model like in its couplings to gauge bosons and the heavy CP even Higgs $H$ having no couplings to gauge bosons. The mass eigenstates $h$ and $H$  in this limit are related to doublet components as follows (see for example ref.~\cite{Gunion:1989we}):
\dis{
\left( 
\begin{array}{c}
h \\
- H \\
\end{array}
\right)
=
\left( 
\begin{array}{cc}
\cos\beta & \sin\beta \\
-\sin\beta & \cos\beta \\
\end{array}
\right)
\left( 
\begin{array}{c}
\sqrt{2} ( {\rm Re} H_d^0 - v_d )\\
 \sqrt{2} ( {\rm Re} H_u^0 - v_u) \\
\end{array}
\right)~.
\label{eq:interactionbasis}
}
The $Y_\nu$ matrix is not proportional to the mass matrix given in eq.~(\ref{eq:mm}) and thus the Higgs couplings are in general flavor violating. Defining couplings of mass eigenstate leptons $f_a$ and $f_b$ to CP-even Higgs bosons  by
\begin{equation}
{\cal L} \supset    - \frac{1}{\sqrt{2}} \, \bar f_{La}  \, \lambda^h_{f_a f_b} \,  f_{Rb}  \, h  - \frac{1}{\sqrt{2}} \, \bar f_{La}  \, \lambda^H_{f_a f_b} \,  f_{Rb}  \, H + h.c.,
\end{equation}
we find:
\dis{
\lambda^h_{\nu_a \nu_b} &= \sin\beta \, (V_L^\dagger Y_\nu V_R)_{ab}~, \\
- \lambda^H_{\nu_a \nu_b} &=  \cos\beta \, (V_L^\dagger Y_\nu V_R)_{ab}~. 
\label{eq:hnunu}
}
Since $Y_\nu v_u = M_\nu - diag (0,M_L,M_N)$, the Higgs boson couplings   can be also written as:
\dis{
\lambda^h_{\nu_a \nu_b} v &= 
\left(
\begin{array}{ccc}
0 & 0 & 0 \\
0 & m_{\nu_4} & 0 \\
0 & 0 & m_{\nu_5} 
\end{array}
\right) 
- V_L^\dagger
\left(
\begin{array}{ccc}
0 & 0 & 0 \\
0 & M_L & 0 \\
0 & 0 & M_N 
\end{array}
\right) 
V_R~, \\
- \lambda^H_{\nu_a \nu_b} v \tan\beta &= 
\left(
\begin{array}{ccc}
0 & 0 & 0 \\
0 & m_{\nu_4} & 0 \\
0 & 0 & m_{\nu_5} 
\end{array}
\right) 
- V_L^\dagger
\left(
\begin{array}{ccc}
0 & 0 & 0 \\
0 & M_L & 0 \\
0 & 0 & M_N 
\end{array}
\right) 
V_R~, 
\label{eq:hnunualt}
}
where we used $v_u = v \sin \beta$. The first terms in above equations represent the expected relations between fermion masses and their couplings to  Higgs bosons and the second term represents contributions from the $M_{L,N}$ terms. This form of couplings makes it obvious that in the absence of vectorlike masses the couplings of $h$ to leptons are fully SM-like, while couplings of $H$ are enhanced by $\tan \beta$ as expected in the limit we assume.

Couplings to charged leptons follow from $H_d^0$ terms in eq.~(\ref{eq:lagrangian}) and can be obtained from those in eqs.~(\ref{eq:hnunu}) with  replacements:  $V_{L,R} \to U_{L,R}$, $Y_\nu \to Y_e$ and  $\beta \to  \beta + \pi/2$, see also ref.~\cite{Dermisek:2013gta} in the case of SM. The corresponding formulas to eqs.~({\ref{eq:hnunualt}}) would show that couplings of $h$ have the usual SM strength, up to contribution from $M_{L,N}$, while couplings of $H$ to charged leptons are suppressed by $\tan \beta$. Finally couplings of the CP-odd Higgs boson, $A$, copy those of $H$ up to the usual $\gamma_5$ factor.

\section{Branching Ratios}
\label{sec:BRs}
We collect expressions for the relevant branching ratios for the process $pp \to H \to \nu_4\nu_\mu\to W\mu \nu_\mu$ and provide several approximate formulas in the limit of small mixing between neutral leptons discussed in the previous section. These formulas will be useful for qualitative understanding of results. From now on, we drop the hat notation for mass eigenstates and also label  the  mass eigenstates $\hat \nu_{L2}$ and $\hat e_2$  as $\nu_\mu$ and $\mu$.

Sizable decay modes of the heavy CP even Higgs boson are $\nu_4 \nu_\mu$, $b\bar b$ and $gg$ for $m_H < 250 \; {\rm GeV}$. As discussed in the previous section we assume that $H$ does not have direct couplings to pairs of gauge bosons and that decay modes to other Higgs bosons are not kinematically possible.  However, our results could be straightforwardly modified to account for additional sizable decay modes of $H$.

The partial decay width of $H \to \nu_4 \nu_\mu$ (where we include both $\bar\nu_4 \nu_\mu$ and $\nu_4 \bar\nu_\mu$ final states) is  given by:
\dis{
\Gamma(H \to \nu_4 \nu_\mu) = \frac{m_H}{8\pi} (\lambda^H_{\nu_\mu \nu_4})^2  \left( 1 - \frac{m_{\nu_4}^2}{m_H^2} \right)^2 ,
\label{eq:gamHtonu4}
}
where  
\begin{eqnarray}
\lambda^H_{\nu_\mu \nu_4} &=&  \cot\beta \left[ \frac{M_L}{v} (V_L^\dagger)_{24} (V_R)_{44} + \frac{M_N}{v} (V_L^\dagger)_{25} (V_R)_{54} \right] \\
&\simeq&  \frac{\kappa_N v \sin\beta  \cos\beta}{M_N}  \, \left[  \bar \kappa +  \frac{M_N(M_L \kappa + M_N \bar \kappa) }{M_N^2  - M_L^2} -  \frac{\bar \kappa (M_L \kappa + M_N \bar \kappa)^2 v^2 \sin^2\beta}{2(M_N^2  - M_L^2)^2} \right],
\label{eq:lam24doublet}
\end{eqnarray}
where the second line is an approximate formula in the limit of small mixing discussed in section~\ref{sec:model}. Note, that this limit assume the $\nu_4$ is mostly the doublet with mass originating from $M_L$. For an approximate formula corresponding to a singlet-like neutral lepton, the  $\lambda^H_{\nu_\mu \nu_5}$ should be used instead. This coupling is given by
\begin{eqnarray}
\lambda_{\nu_\mu \nu_5}^H &=& \cot\beta \left[ \frac{M_L}{v} (V_L^\dagger)_{24} (V_R)_{45} + \frac{M_N}{v} (V_L^\dagger)_{25} (V_R)_{55} \right] \\
&\simeq&   - \kappa_N \cos\beta\left[   1 - \frac{(M_L \kappa + M_N \bar \kappa)^2v^2 \sin^2\beta }{2(M_N^2 - M_L^2)^2}  - \frac{\bar \kappa v^2 \sin^2\beta}{M_N} \frac{M_L \kappa + M_N \bar \kappa}{M_N^2 - M_L^2} \right],
\label{eq:lam24singlet}
\end{eqnarray}
where the second line is an appropriate approximate formula in the case of singlet-like lepton with mass originating from $M_N$.

The decay width of $H \to b \bar b$ is given by
\dis{
\Gamma(H \to b \bar b) &= \frac{3 G_F}{4\sqrt{2} \pi} m_H \bar m_b^2 (m_H) \tan^2\beta \left[ 1 + \Delta_{qq} + \Delta_H^2 \right],
}
where $\bar m_b (m_H)$ is the running b-quark mass evaluated at the scale $m_H$ and the  correction factors $\Delta_{qq}$ and $\Delta_H^2$ can be found in ref.~\cite{Djouadi:2005gi}. The decay width of $H \to gg$ is given by
\dis{
\Gamma(H \to g g) &= \frac{G_F \alpha_S^2 m_H^3 \cot^2\beta}{36 \sqrt{2} \pi^3} \left| \frac34 A_{1/2} \right|^2~,
}
with
\dis{
A_{1/2} = 2\frac{\tau + (\tau - 1) f(\tau)}{\tau^2}~,
}
where $\tau = m_H^2 / 4 m_t^2$ and $f(\tau) = \arcsin^2 \sqrt{\tau}$  for $\tau \le 1$. 

The branching ratio of $H \to \nu_4 \nu_\mu$ is then given by 
\dis{
{\rm BR}(H \to \nu_4 \nu_\mu) = \frac{\Gamma(H \to \nu_4 \nu_\mu)}{\Gamma(H \to \nu_4 \nu_\mu) + \Gamma(H \to b \bar b) + \Gamma(H \to g g)}~.
}

The neutral lepton $\nu_4$ can decay into standard model leptons and the Higgs, $W$, and $Z$ bosons. 
Neglecting the muon mass, the partial decay width of $\nu_4 \to h \nu_\mu$ is  given by:
\dis{
\Gamma(\nu_4 \to h \nu_\mu) = \frac{m_{\nu_4}}{16\pi} (\lambda^h_{\nu_\mu \nu_4})^2  \left( 1 - \frac{m_h^2 }{m_{\nu_4}^2} \right)^2 \; ,
}
where  
\dis{
\lambda^h_{\nu_\mu \nu_4} =   -\frac{M_L}{v} (V_L^\dagger)_{24} (V_R)_{44} - \frac{M_N}{v} (V_L^\dagger)_{25} (V_R)_{54} .
}
The partial decay width of $\nu_4 \to W \mu$ is  given by:
\dis{
\Gamma(\nu_4 \to W^+ \mu^-) = \frac{m_{\nu_4}}{32 \pi} \left[(g_L^{W \nu_4 \mu})^2 + (g_R^{W \nu_4 \mu})^2 \right] \frac{m_{\nu_4}^2}{M_W^2} \left( 1 - \frac{M_W^2}{m_{\nu_4}^2} \right)^2 \left( 1 + 2\frac{M_W^2}{m_{\nu_4}^2} \right) \; ,
}
where  
\begin{eqnarray}
g_L^{W \nu_4 \mu} &=&  \frac{g}{\sqrt{2}}\left[ (V_L^\dagger)_{42} (U_L)_{22} + (V_L^\dagger)_{44} (U_L)_{42} \right] \; ,  \label{Wcoupling}\\
g_R^{W \nu_4 \mu} &=& \frac{g}{\sqrt{2}} (V_R^\dagger)_{44} (U_R)_{42} \; .
\end{eqnarray}
Finally, the partial decay width of $\nu_4 \to Z \nu_\mu$ is  given by:
\dis{
\Gamma(\nu_4 \to Z \nu_\mu) &= \frac{m_{\nu_4}}{32 \pi}  (g_L^{Z \nu_\mu \nu_4})^2 \frac{m_{\nu_4}^2}{M_Z^2} \left( 1 - \frac{M_Z^2}{m_{\nu_4}^2} \right)^2 \left( 1 + 2\frac{M_Z^2}{m_{\nu_4}^2} \right) \; ,
}
where
\dis{
g_L^{Z \nu_\mu \nu_4} =   \frac{g}{2 \cos\theta_W}\left[(V_L^\dagger)_{42} (V_L)_{22} + (V_L^\dagger)_{44} (V_L)_{42} \right].
\label{Zcoupling}
}

Assuming only these decay modes of $\nu_4$, the branching ratio of $\nu_4 \to W \mu$  is given by
\dis{
{\rm BR}(\nu_4 \to W^+ \mu^-) &= \frac{\Gamma(\nu_4 \to W^+ \mu^-)}{\Gamma(\nu_4 \to h \nu_\mu) + \Gamma(\nu_4 \to W^+ \mu^-) + \Gamma(\nu_4 \to Z \nu_\mu)}~.
}
The branching ratio of $H\to W\mu\nu_\mu$  is defined as 
\begin{equation}
{\rm BR}(H\to W\mu\nu_\mu) \; = \; {\rm BR}(H \to \nu_4 \nu_\mu) \; {\rm BR}(\nu_4 \to W^+ \mu^-).
\end{equation}

\section{Parameter space scan and constraints from precision electroweak data and direct production}
\label{sec:constraints}
We perform a scan over all the model parameters introduced in section~\ref{sec:model} over the ranges
\begin{align}
M_{L,N} &\in [0, 500] {\rm GeV} \; , \\
\kappa_N, \kappa, \bar \kappa &\in [-0.5, 0.5] \; , \\
\tan\beta & \in [0.3, 3] \; . 
\end{align}
We fix the mass term of the $SU(2)$ singlet charged vectorlike lepton $M_E = 1000~{\rm GeV}$. We simplify the decay patterns of the heavy Higgs by requiring $m_{\nu_5} > m_H$ (to avoid $H\to \nu_5 X$ channels) and $m_{\nu_4} > m_H/2$ (to avoid decays into pairs of heavy vectorlike leptons). Moreover we include mixing exclusively in the neutral sector. 

We impose constraints from precision EW data related to the muon and muon neutrino that include the $Z$ pole observables ($Z$ partial width to $\mu^+\mu^-$, the invisible width, forward-backward asymmetry, left-right asymmetry), the $W$  partial width, and the muon lifetime. We also impose constraints from oblique corrections, namely from S and T parameters. Unless specified otherwise these are obtained from~ref.~\cite{Agashe:2014kda}. Finally, we impose limits from direct searches: the LEP limits on masses of  new charged leptons, 105 GeV, and the limits on pair production of vectorlike leptons at the LHC summarized in ref.~\cite{Dermisek:2014qca}. Constraints on the production of heavy Higgs will be discussed in the following section together with results.

Constraints on the muon couplings were already discussed in ref.~\cite{Dermisek:2013gta}. Precision electroweak measurements constrain modification of couplings of the muon to the $Z $ and $W$ bosons at $\sim 0.1$\% level which, in the limit of small mixing,   approximately translates into 95\% C.L. bounds on $\lambda_{E,L}$ couplings:
\begin{equation}
\left|\frac{\lambda_E v_d}{M_E} \right| \lesssim 0.03, \quad \quad \left|\frac{\lambda_L v_d}{M_L} \right| \lesssim 0.04,
 \label{eq:lambda_ELmax}
 \end{equation}
assuming only mixing (Yukawa couplings) in the charged sector.

In the neutral lepton sector the strongest limits are obtained  from  the muon lifetime. In what follows we discuss this limit together with  the invisible widths of the Z boson and constraints from direct production of vectorlike leptons.

\subsection{The muon lifetime}
\label{sec:muonlifetime}
The Fermi constant $G_F $ is determined with a high precision  from  the measurement of   muon lifetime. In the standard model $G_F = (\sqrt{2}/8) g^2 / M_W^2$ while in our model one of the $g/\sqrt{2}$ factors is replaced by $g_L^{W \nu_\mu \mu}$  given by
\dis{
g_L^{W \nu_\mu \mu} = \frac{g}{\sqrt{2}} \left[ (V_L^\dagger)_{22} (U_L)_{22} + (V_L^\dagger)_{24} (U_L)_{42} \right].
\label{eq:muontau}  
}
The allowed range for $g_L^{W \nu_\mu \mu}$ is obtained from the uncertainty in the $W$ mass, $M_W = 80.385 \pm 0.015$ GeV. The relative uncertainty in $M_W^2$ is   $2 \times \frac{0.015}{80.385} = 3.73 \times 10^{-4}$ and we set the 95\% C.L.  upper limit on the deviation of $g_L^{W \nu_\mu \mu}$ from $g/\sqrt{2}$ as:
\dis{
\left|\frac{g_L^{W \nu_\mu \mu}}{g/\sqrt{2}} - 1\right| < 2 \times 3.73 \times 10^{-4} = 7.46 \times 10^{-4}~.  
}
Assuming no mixing in the charged sector and using the small mixing approximation eq.~(\ref{eq:VL}),
\dis{
g_L^{W \nu_\mu \mu} = \frac{g}{\sqrt{2}} (V_L^\dagger)_{22} \simeq  \frac{g}{\sqrt{2}} \left(  1  - \frac{\kappa_N^2 v_u^2}{2M_N^2}  \right)~,
\label{eq:gLW_app} 
}
and we obtain an approximate 95\% C.L. upper bound on  the size of $\kappa_N$ coupling:
\begin{equation}
\left|\frac{\kappa_N v_u}{M_N} \right| \lesssim 0.04.
\label{eq:kappa_N_limit}
\end{equation}
Considering also mixing in the charged sector, the bound is shared between $\kappa_N$ and $\lambda_E$:
\begin{equation}
\sqrt{\left(\frac{\kappa_N v_u}{M_N}\right)^2 + \left(\frac{\lambda_E v_d}{M_E} \right)^2}\lesssim 0.04.
\end{equation}

The partial decay width of  $W \to \mu \nu_\mu$ depends quadratically on the $g_L^{W \nu_\mu \mu}$ coupling. However, it is measured with about 2\% precision and thus the resulting constraint on  the coupling is significantly weaker.

\subsection{Invisible width of Z}
\label{sec:Zinv}
\begin{figure}
\begin{center}
\includegraphics[width=.49\linewidth]{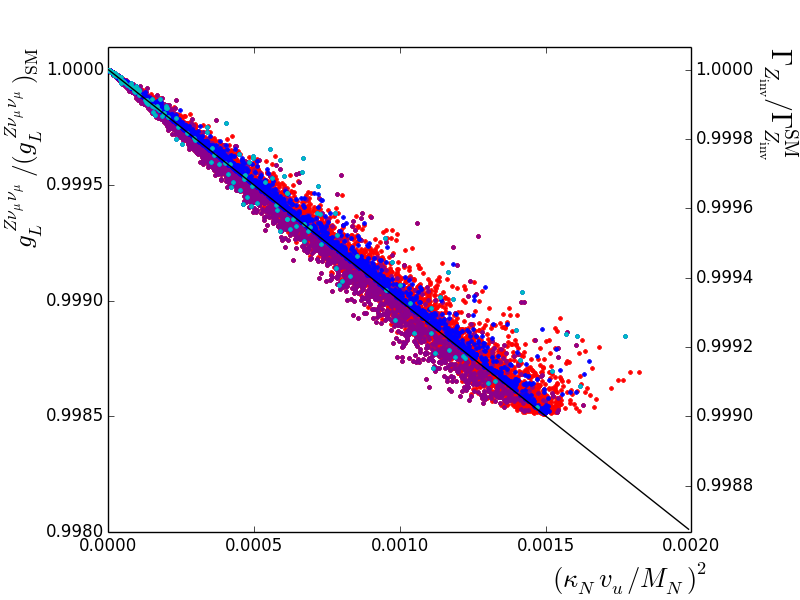}
\includegraphics[width=.49\linewidth]{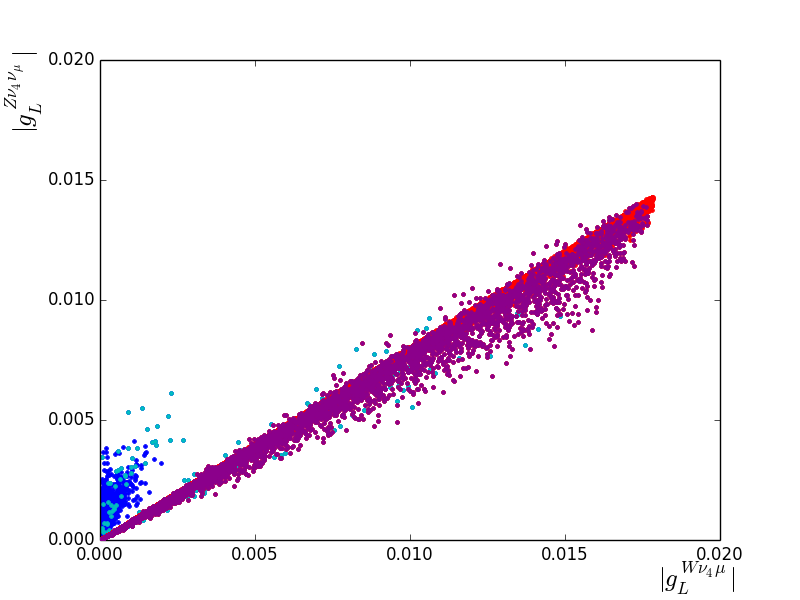}
\caption{ Parameter space scan points (for $m_H = 155$ GeV and $m_{\nu_4} = 135$ GeV) that survive all EW precision constraints. The blue, cyan, magenta and red points have singlet fraction (defined as $| (V_L^\dagger)_{45}|^2 / 2 + | (V_R^\dagger)_{45} |^2 /2$) in the ranges [0,5]\%, [5,50]\%, 
[50,95]\%, and [95,100]\%, respectively. In the left panel, we show the $(\kappa_N v_u/M_N)^2 - g_L^{Z \nu_\mu \nu_\mu}/(g_L^{Z \nu_\mu \nu_\mu})_{\rm SM}$ plane; on the right axis we show the corresponding $\Gamma_{Z_{\rm inv}}^{\rm exp}/\Gamma_{Z_{\rm inv}}^{\rm SM}$ ratio. In the right panel we show the off-diagonal $|g_L^{W\nu_4\mu}|$ and $|g_L^{Z\nu_4\nu_\mu}|$ gauge couplings.}
\label{fig:Zinv}
\end{center}
\end{figure}

The partial width of $Z \to \nu_\mu \bar \nu_\mu$ is given by
\dis{
\Gamma(Z \to \nu_\mu \bar \nu_\mu) &= \frac{M_Z}{24\pi} (g_L^{Z \nu_\mu \nu_\mu})^2 
}
where
\begin{eqnarray}
g_L^{Z \nu_\mu \nu_\mu} &=& \frac{g}{2 \cos \theta_W} \left[ (V_L^\dagger)_{22} (V_L)_{22} + (V_L^\dagger)_{24} (V_L)_{42} \right] \\
&\simeq& \frac{g}{2 \cos \theta_W} \left(   1- \frac{\kappa_N^2 v_u^2}{M_N^2}  \right),
\end{eqnarray}
where the second line is an appropriate approximate formula in the case of small mixing. In this limit, the upper bound on $\kappa_N$ obtained  from  muon lifetime, eq.~(\ref{eq:kappa_N_limit}), suggests that $\Gamma(Z \to \nu_\mu \bar \nu_\mu)$ can be modified at most at 0.3\% level (the sign of the correction is always negative). Since we assume that only one generation of SM leptons mix with vectorlike pairs, the invisible width of Z can be lowered at most by 0.1\%.  This is also visible in figure~\ref{fig:Zinv} where we consider randomly generated points in the $\kappa_N$, $\kappa$, $\bar\kappa$, $M_L$ and $M_N$  parameter space for fixed $m_H = 155$ GeV and $m_{\nu_4} = 135$ GeV (different choices of masses do not sizably affect the allowed ranges), assuming no mixing in the charged lepton sector, and impose all EW precision constraints (including direct productions bounds discussed in section~\ref{sec:direct} below). In the left and right panels of figure~\ref{fig:Zinv} we consider the $(\kappa_N v_u/M_N)^2 - g_L^{Z \nu_\mu \nu_\mu}/(g_L^{Z \nu_\mu \nu_\mu})_{\rm SM}$,  $(\kappa_N v_u/M_N)^2 - \Gamma_{Z_{\rm inv}}^{\rm exp}/\Gamma_{Z_{\rm inv}}^{\rm SM}$ and $|g_L^{W\nu_4\mu}| - |g_L^{Z\nu_4\nu_\mu}|$ planes. Both the upper limit on $\kappa_N v_u/M_N$,  eq.~(\ref{eq:kappa_N_limit}), and the resulting largest possible effect in $\Gamma_{Z_{\rm inv}}$ follow closely those obtained from the approximate formulas. Points with $\nu_4$ being mostly singlet (red points) or mostly doublet (blue points)  cluster very near the line that assumes the approximate relation in eq.~(\ref{eq:gLW_app}) is exact and even highly mixed scenarios (cyan and magenta) are not very far.

Since the ratio of the measured value of the $Z$-boson invisible decay width and its  SM expectation is
\dis{
\frac{\Gamma_{Z_{\rm inv}}^{\rm exp}}{\Gamma_{Z_{\rm inv}}^{\rm SM}} = \frac{499.0 \pm 1.5 ~({\rm MeV})}{501.66 \pm 0.05 ~({\rm MeV})} = 0.995 \pm 0.003~,
}
the invisible width does not provide additional constraint to that obtained from the muon lifetime. This however assumes that only the 2nd generation of SM leptons mixes with vectorlike leptons. The effect on invisible width can be larger if more generations mix with vectorlike leptons or if one considers mixing with the 3rd generation instead of the 2nd since the constraints on 3rd generation couplings are weaker.

In conclusion, couplings of SM gauge bosons to the second family of leptons, $g_L^{W \nu_\mu \mu}$ and $g_L^{Z \nu_\mu \nu_\mu}$, can deviate from their SM values by less than $\sim 0.1$\%. Moreover, within the explicit model we consider, these constraints imply upper limits of order $\sim 0.02$ on the new off-diagonal $g_L^{W\nu_4\mu}$ and $g_L^{Z\nu_4\nu_\mu}$ gauge couplings as we can see in the right panel of figure~\ref{fig:Zinv}.

\subsection{Direct production of vectorlike leptons}
\label{sec:direct}
Let us first consider constraints from Drell-Yan production of $\nu_4 \nu_4$ or $\nu_4 e_4$ leading to  at least 3 leptons  in the final state and  $E_T^{\rm miss}$. The cross section of  $p p \to \nu_4 \nu_4$ is proportional to $(g_L^{Z \nu_4 \nu_4})^2 + (g_R^{Z \nu_4 \nu_4})^2$ where the couplings are given in eqs.~(\ref{eq:gZL})--(\ref{eq:gZR}). Thus the cross section is modified from the one that corresponds to fully doublet $\nu_4$  by factor 
\begin{align}
R_{\nu_4 \nu_4} \equiv \frac{1}{2} \left[ (V_L^\dagger)_{42} (V_L)_{24} + (V_L^\dagger)_{44} (V_L)_{44} \right]^2 + \frac{1}{2} \left[ (V_R^\dagger)_{44} (V_R)_{44} \right]^2 ~.
\label{eq:rnn}
\end{align}
At present, searches for anomalous production of multilepton events constrain only the case when both $\nu_4$s decay to $W \mu$ and the limits on $R_{\nu_4 \nu_4} \times$ BR$^2 (\nu_4 \to W \mu)$ can be read from Table 2 of ref.~\cite{Dermisek:2014qca}. They are summarized in our Table~\ref{table:dynn} for  reference values of $m_{\nu_4}$. In our numerical analysis we interpolate these results for other values of $m_{\nu_4}$.

\begin{table}[t]
\begin{center}
\begin{tabular}{c c c c c c}
\hline
\hline
$m_{\nu_4}$ [GeV] & ~~105 & ~~125 & ~~150 & ~~200 & ~~300 \\
\hline
$[R_{\nu_4 \nu_4} \times$ BR$^2 (\nu_4 \to W \mu)]_{max}$ & ~~0.090 & ~~0.141 & ~~0.141 & ~~0.164 & ~~0.582 \\
\hline
$[R_{e_4 \nu_4} \times$  BR$(\nu_4 \to W \mu)]_{max}$ & ~~0.109 & ~~0.203 & ~~0.267 & ~~0.355 & ~~1 \\

\hline
\hline
\end{tabular}
\end{center}
\caption{Upper bounds on $R_{\nu_4 \nu_4} \times$ BR$^2 (\nu_4 \to W \mu)$ and $R_{e_4 \nu_4} \times$  BR$(\nu_4 \to W \mu)$  obtained from ref.~\cite{Dermisek:2014qca} for several masses of $\nu_4$. The limits on $R_{e_4 \nu_4} \times$  BR$(\nu_4 \to W \mu)$ assume that BR$(e_4 \to W \nu_\mu) = 1$  and $m_{e_4} = m_{\nu_4}$. \label{table:dynn}
}
\end{table}
\begin{figure}
\begin{center}
\includegraphics[width=.55\linewidth]{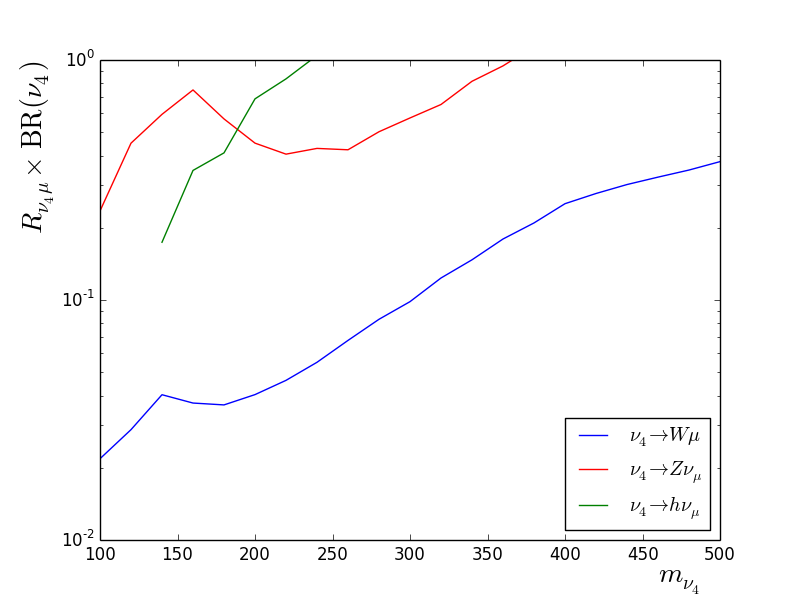}
\caption{Upper bounds on $R_{\nu_4 \mu} \times$ BR($\nu_4 \to W \mu, Z \nu_\mu, h \nu_\mu$) as functions of $m_{\nu_4}$. 
}
\label{fig:singlenu4}
\end{center}
\end{figure}

Similarly, the cross section of  $p p \to e_4 \nu_4$  is proportional to $(g_L^{W \nu_4 e_4})^2 + (g_R^{W \nu_4 e_4})^2$ where the couplings are given in eqs.~(\ref{eq:gWL})--(\ref{eq:gWR}). Thus the cross section is modified from the one that corresponds to fully doublet $\nu_4$ and $e_4$ by factor 
\dis{
R_{e_4 \nu_4} \equiv \frac{1}{2} \left[   (V_L^\dagger)_{42} (U_L)_{24} + (V_L^\dagger)_{44} (U_L)_{44} \right]^2 
+ \frac{1}{2} \left[ (V_R^\dagger)_{44} (U_R)_{44} \right]^2~.
\label{eq:ren}
}
The analysis in ref.~\cite{Dermisek:2014qca}, assuming  $m_{e_4} = m_{\nu_4}$, shows strong limits on this production cross section. Some decay modes  of $e_4$ and $\nu_4$ are consistent with data only for  masses higher than 500 GeV. In our analysis we are focusing on the case with no mixing in the charged lepton sector. In this case BR$(e_4 \to W \nu_\mu) =1 $ (the required coupling originates from mixing in the neutral sector) and the limits on  $R_{e_4 \nu_4} \times$  BR$(\nu_4 \to W \mu)$ can be obtained from Table 2 of \cite{Dermisek:2014qca}. Other decay modes of $\nu_4$ are not constrained in this case. The upper bounds are summarized in Table~\ref{table:dynn}. We again interpolate these results for other values of $m_{\nu_4}$.

Finally, let us comment on a single production of a new lepton.  The $W - \nu_4 - \mu$ coupling results in a production of $\nu_4$ through the process $ p p \to W^\ast \to \nu_4 \mu$ which also can lead to at least 3 leptons with $E_T^{\rm miss}$ in the final state. The bounds  were discussed in ref.~\cite{Das:2014jxa} in the context of  TeV scale seesaw models with very small lepton number violating terms,  see for example  ref.~\cite{BhupalDev:2012zg}, which are not constrained from the same-sign dilepton searches.
 
 The cross section of  $p p \to \nu_4 \mu$  is proportional to $(g_L^{W \nu_4 \mu})^2 + (g_R^{W \nu_4 \mu})^2$ where the couplings are given in eqs.~(\ref{eq:gWL})--(\ref{eq:gWR}). Thus the cross section is modified from the one that corresponds to full strength coupling of the two leptons to $W$ by factor 
\dis{
R_{\nu_4 \mu} \equiv \left[(V_L^\dagger)_{42} (U_L)_{22} + (V_L^\dagger)_{44} (U_L)_{42} \right]^2 + \left[ (V_R^\dagger)_{44} (U_R)_{42} \right]^2~.
}
We closely follow ref.~\cite{Dermisek:2014qca} to set limits from the ATLAS searches for anomalous production of multilepton events~\cite{TheATLAScollaboration:2013cia} on three decay modes:
\dis{
p p \to W^\ast \to \nu_4 \mu \to W \mu \mu~,  ~ Z \nu_\mu \mu~, ~ h \nu_\mu \mu~,
}
where $h$ is the SM Higgs with mass 125 GeV. The obtained upper bounds on $R_{\nu_4 \mu} \times$ BR($\nu_4 \to W \mu, Z \nu_\mu, h \nu_\mu$) are shown in Fig. \ref{fig:singlenu4} as functions of $m_{\nu_4}$.  We see that for any combination of branching ratios the constraint on $R_{\nu_4 \mu}$ is  at most of $\mathcal{O}(10^{-2})$.  This limit is much weaker than those obtained from precision EW data; in fact, for the surviving points in figure~\ref{fig:Zinv} the maximum value of $R_{\nu_4 \mu}$ is of $\mathcal{O}( 10^{-3})$.

\section{Main results: contributions of $H\to W\mu\nu$ to $pp \to WW$ and $H \to WW$ }
\label{sec:results}
\begin{figure}[h]
\begin{center}
\includegraphics[width=.46\linewidth]{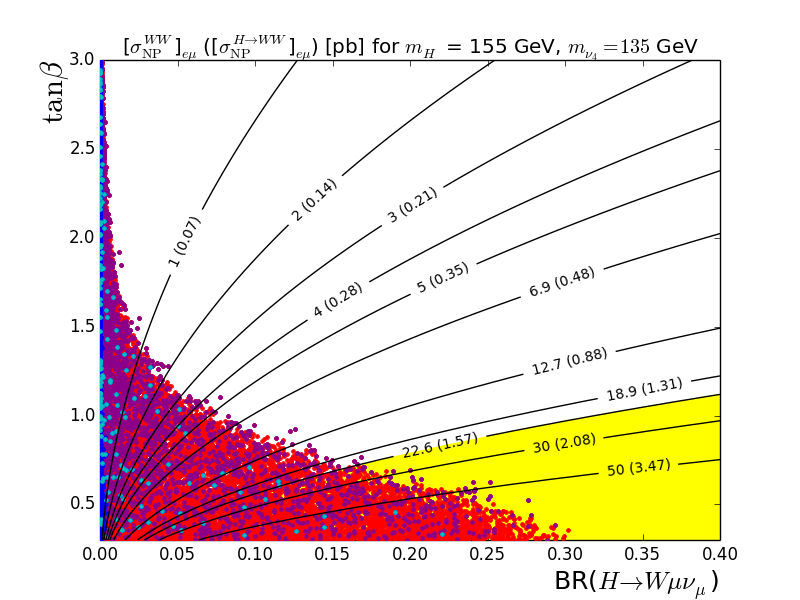}
\includegraphics[width=.46\linewidth]{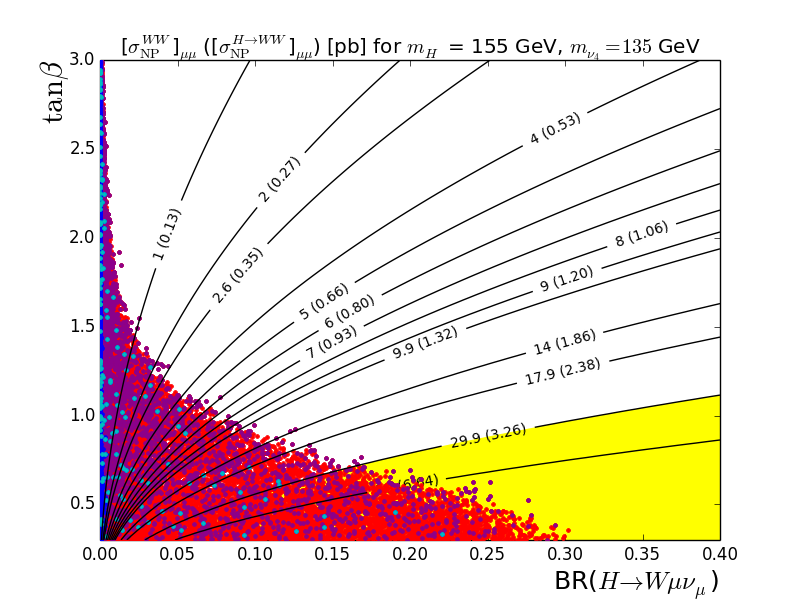}
\includegraphics[width=.46\linewidth]{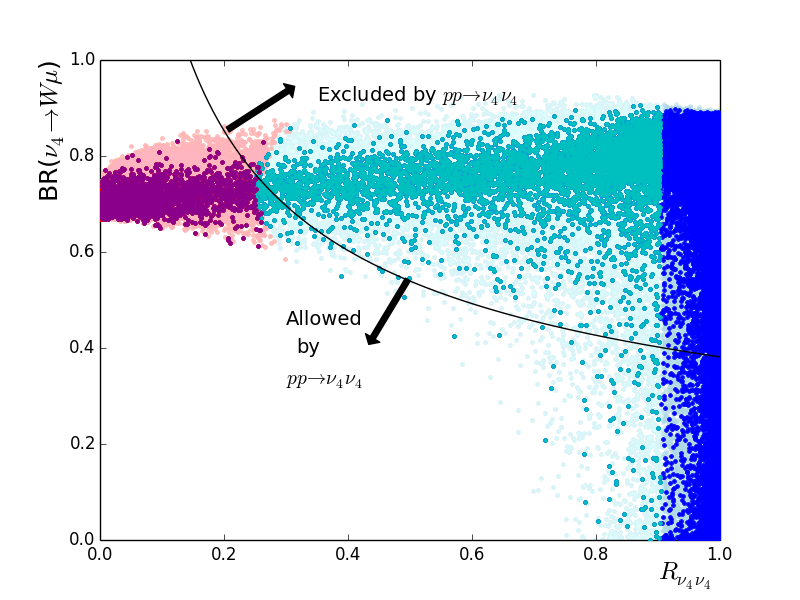}
\includegraphics[width=.46\linewidth]{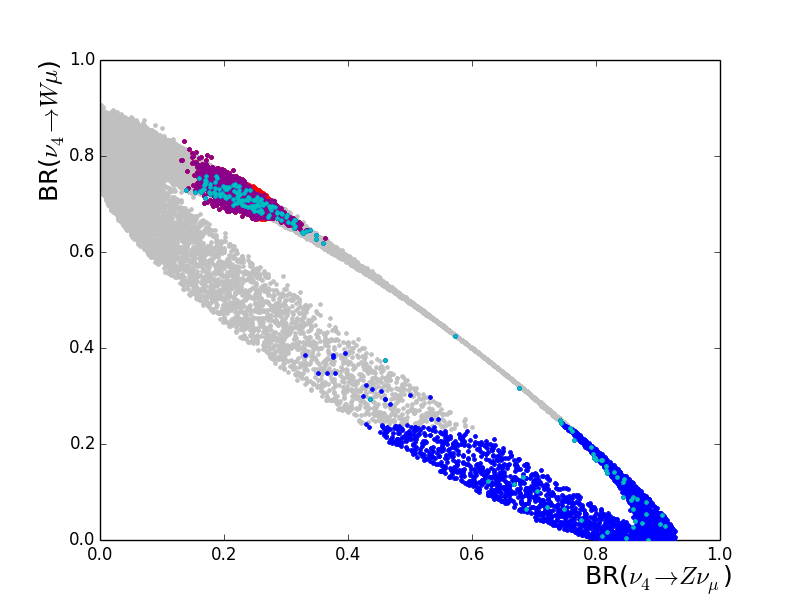}
\includegraphics[width=.46\linewidth]{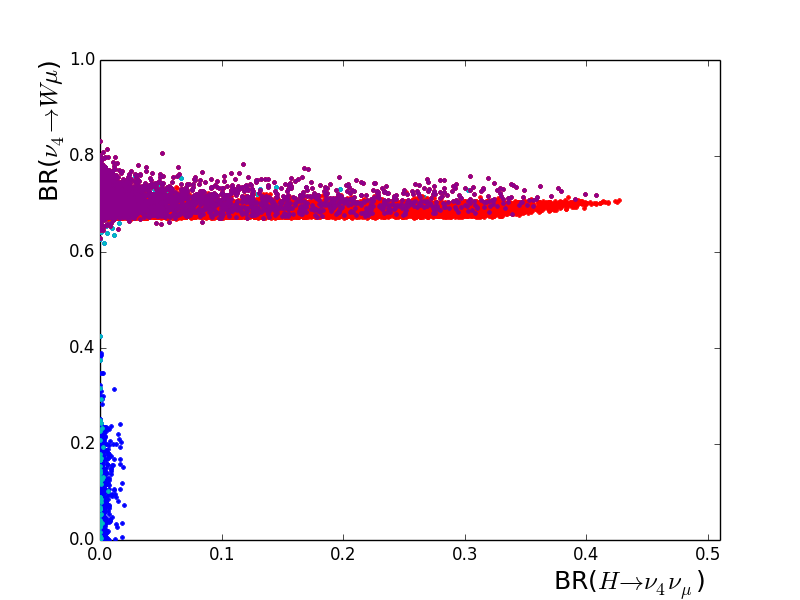}
\includegraphics[width=.46\linewidth]{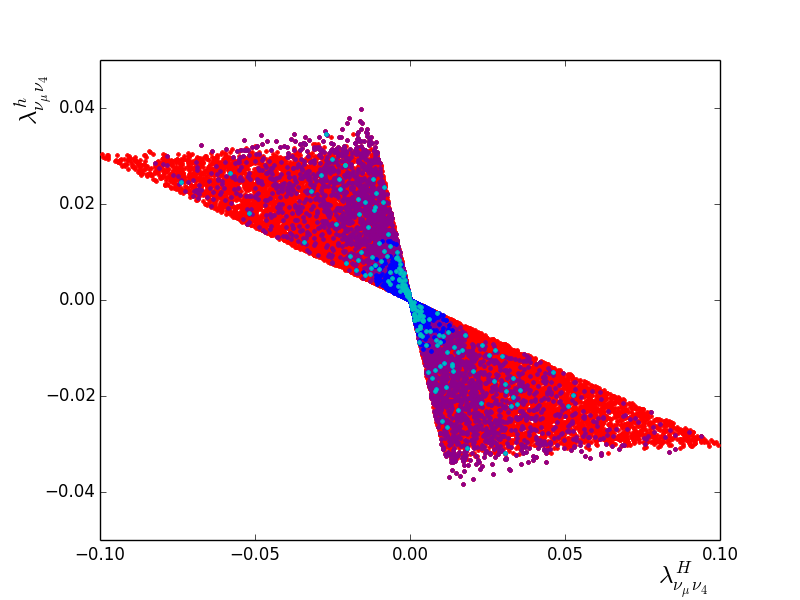}
\caption{Parameter space scan for $m_H = 155\; {\rm GeV}$ and $m_{\nu_4} = 135\; {\rm GeV}$. The blue, cyan, magenta and red points have singlet fraction in the ranges [0,5]\%, [5,50]\%, [50,95]\%, and [95,100]\%, respectively. In the two upper plots all constraints are imposed and we focus on the ${\rm BR}(H \to W\mu \nu_\mu) - \tan\beta$ plane for the $e\mu \nu_e \nu_\mu$ and $\mu\mu \nu_\mu \nu_\mu$ final states, respectively. The black contours are the values of the effective $pp\to WW$ and $pp\to H\to WW$ cross sections (in pb) defined in eq.~(\ref{effectiveXS}). The yellow shaded area is excluded by $H\to WW$ searches. In the middle-left plot, we consider the $R_{\nu_4\nu_4} - {\rm BR}(\nu_4 \to W\mu )$. Here the light-shaded points do not satisfy the muon lifetime constraint and the impact of multilepton + $E_T^{\rm miss}$ searches from Drell-Yan pair production process $pp \to \nu_4 \bar \nu_4 \to W^+ W^- \mu^+ \mu^-$ is indicated by the black curve. In the middle-right plot we show the ${\rm BR}(\nu_4\to Z \nu_\mu) - {\rm BR}(\nu_4 \to W\mu )$. Here the gray points are excluded by multilepton searches. In the two lower plots we consider the ${\rm BR}(H\to \nu_4 \nu_\mu) - {\rm BR}(\nu_4 \to W\mu )$ and $\lambda_{\nu_\mu \nu_4}^H - \lambda_{\nu_\mu \nu_4}^h$ planes.
\label{fig:155135}}
\end{center}
\end{figure}

\begin{figure}
\begin{center}
\includegraphics[width=.46\linewidth]{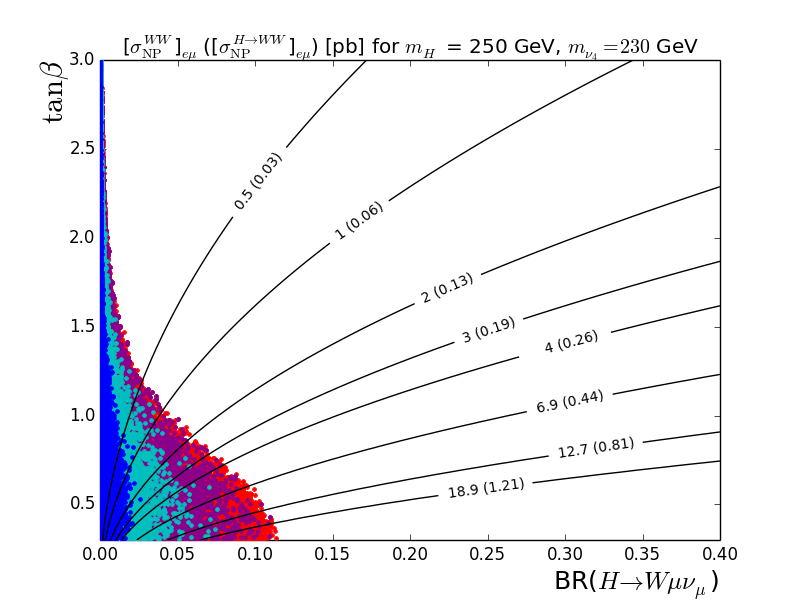}
\includegraphics[width=.46\linewidth]{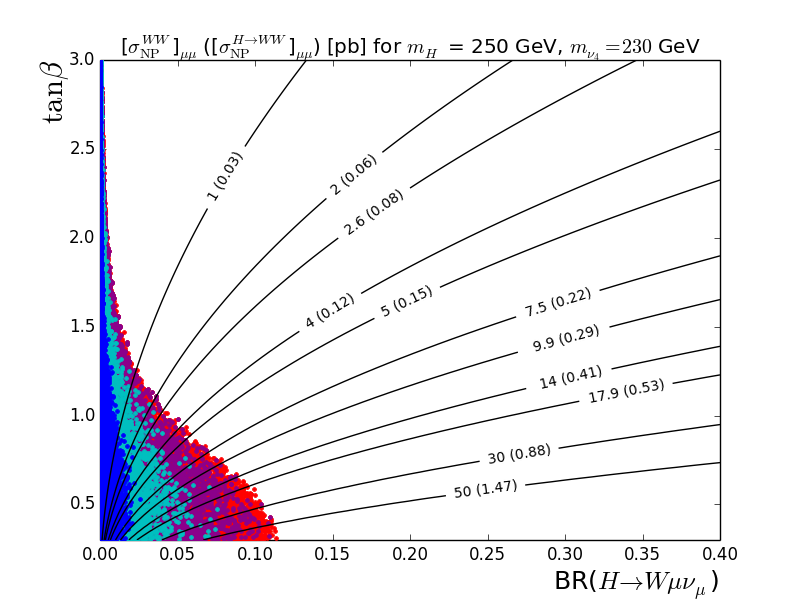}
\includegraphics[width=.46\linewidth]{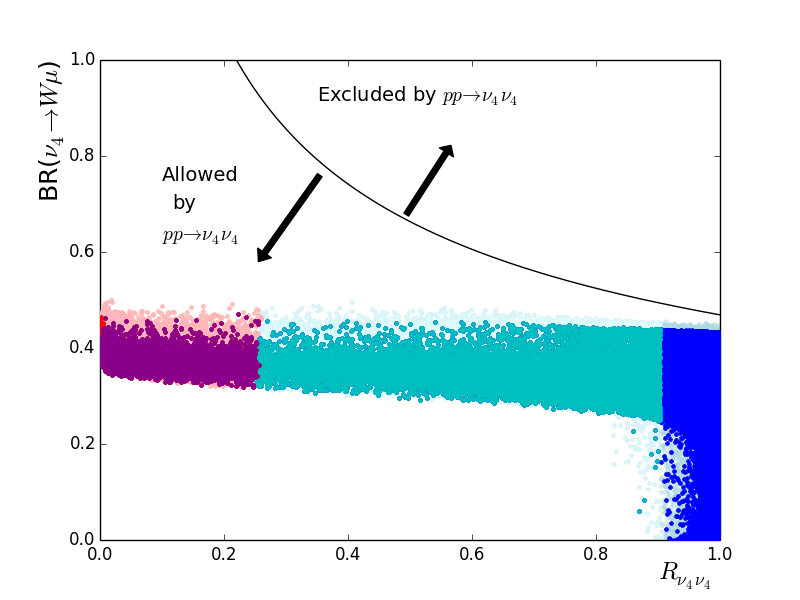}
\includegraphics[width=.46\linewidth]{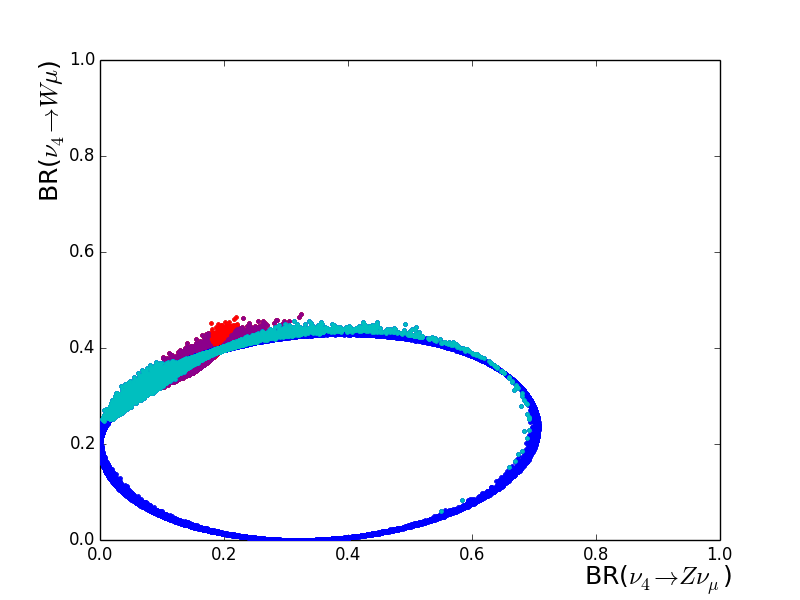}
\includegraphics[width=.46\linewidth]{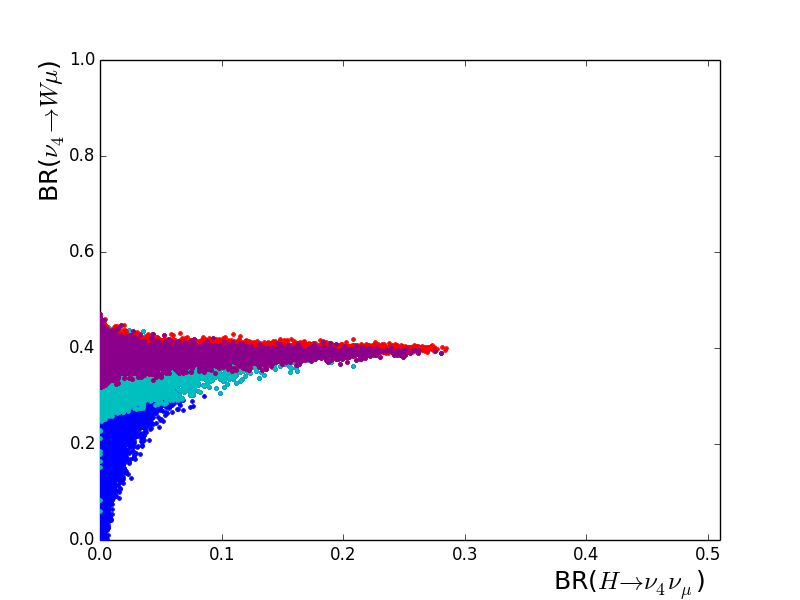}
\includegraphics[width=.46\linewidth]{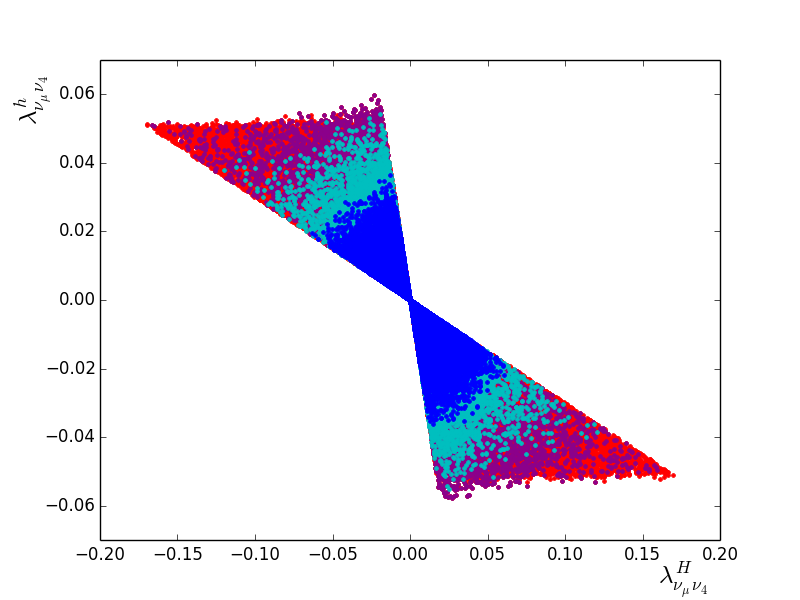}
\caption{Parameter space scan for $m_H = 250$ GeV and $m_{\nu_4} = 230$ GeV. See the caption in figure~\ref{fig:155135} for further details. \label{fig:250230}}
\end{center}
\end{figure}

\begin{figure}
\begin{center}
\includegraphics[width=.49\linewidth]{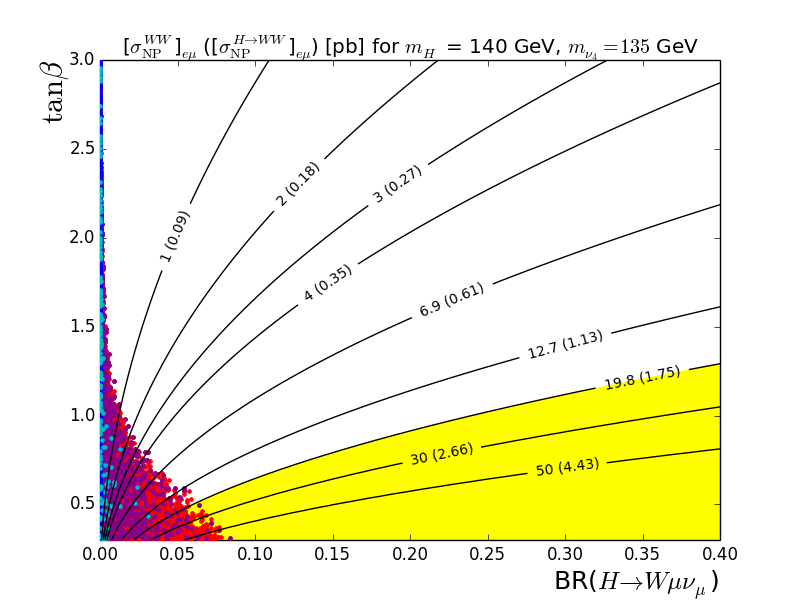}
\includegraphics[width=.49\linewidth]{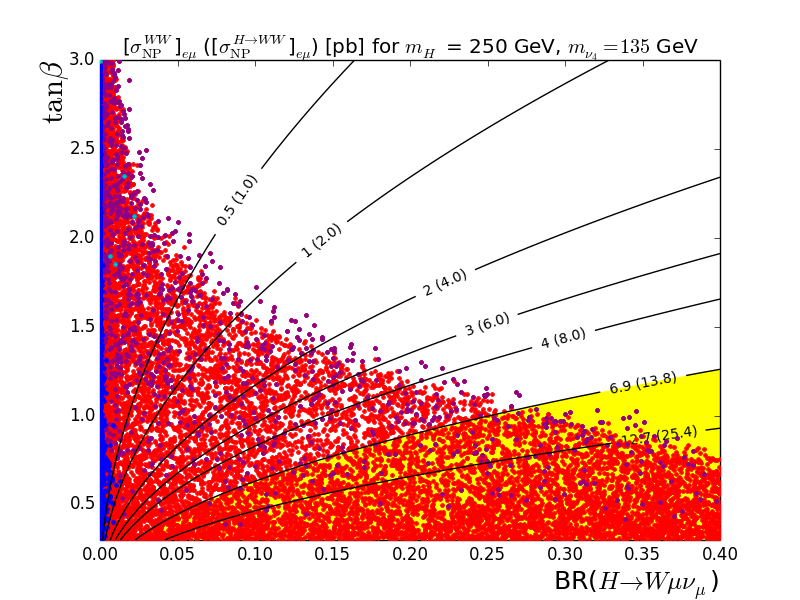}
\includegraphics[width=.49\linewidth]{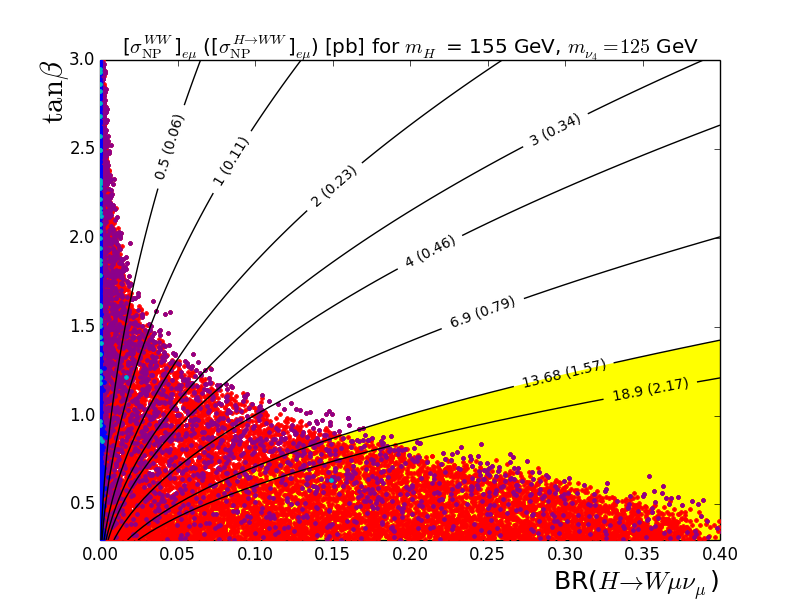}
\includegraphics[width=.49\linewidth]{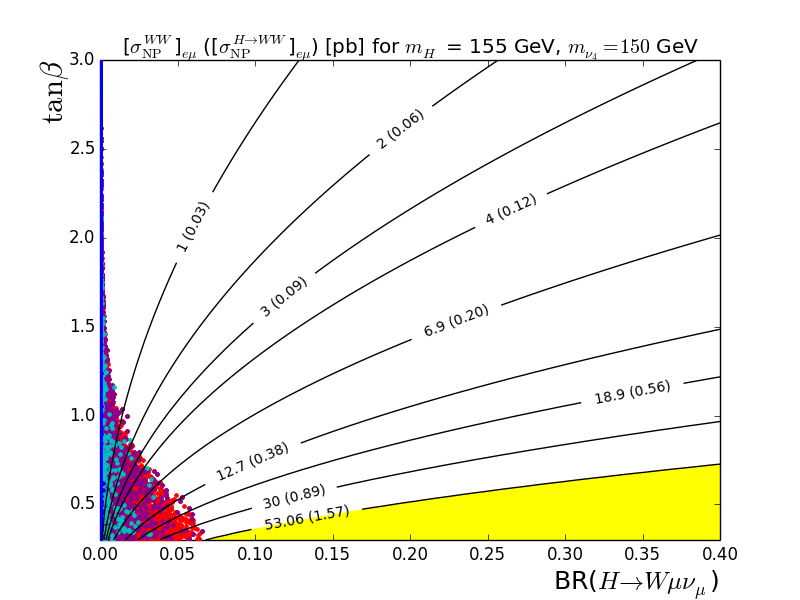}
\caption{Projection onto the $[{\rm BR}(H \to W\mu \nu_\mu),\tan\beta]$ plane of the parameter scan described in the main text for various values of $m_H$ and $m_{\nu_4}$. The black contours are the values of the effective $pp\to WW$ and $pp\to H\to WW$ cross sections (in pb) for the $e\mu \nu_e \nu_\mu$ final state. See the caption in figure~\ref{fig:155135} for further details. \label{fig:mHmNemu}}
\end{center}
\end{figure}

\begin{figure}
\begin{center}
\includegraphics[width=.49\linewidth]{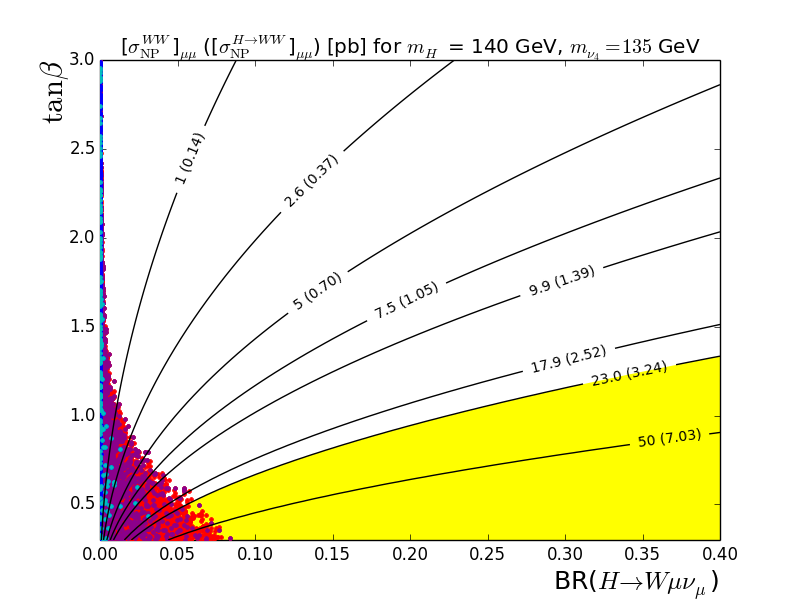}
\includegraphics[width=.49\linewidth]{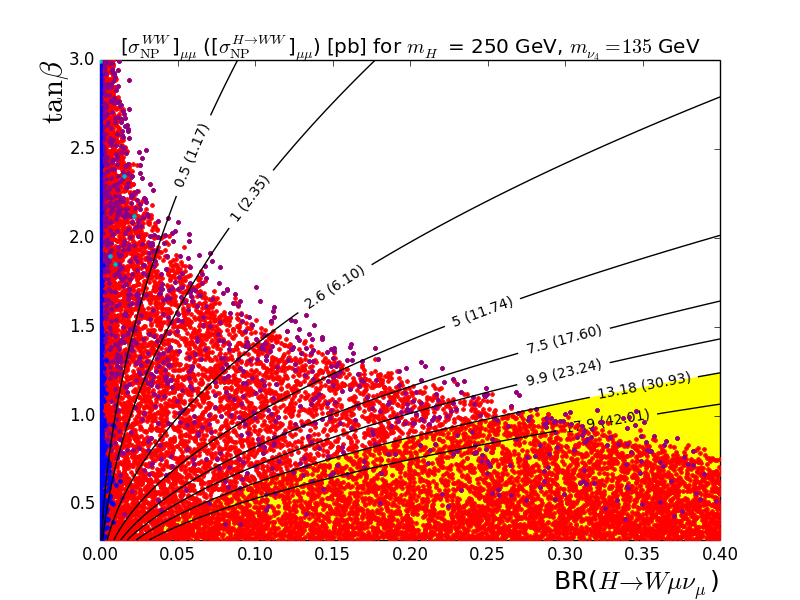}
\includegraphics[width=.49\linewidth]{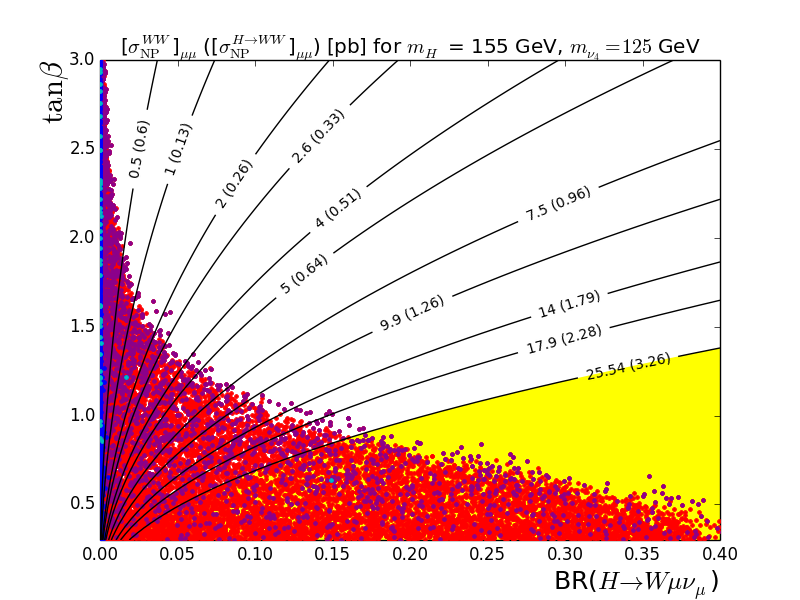}
\includegraphics[width=.49\linewidth]{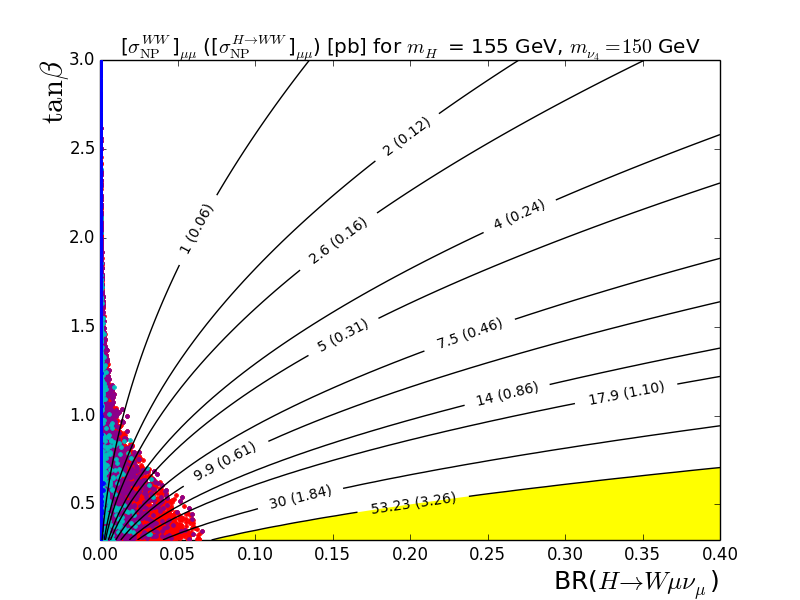}
\caption{Projection onto the $[{\rm BR}(H \to W\mu \nu_\mu),\tan\beta]$ plane of the parameter scan described in the main text for various values of $m_H$ and $m_{\nu_4}$. The black contours are the values of the effective $pp\to WW$ and $pp\to H\to WW$ cross sections (in pb) for the $\mu\mu \nu_\mu \nu_\mu$ final state. See the caption in figure~\ref{fig:155135} for further details. \label{fig:mHmNmumu}}
\end{center}
\end{figure}

\begin{figure}
\begin{center}
\includegraphics[width=.49\linewidth]{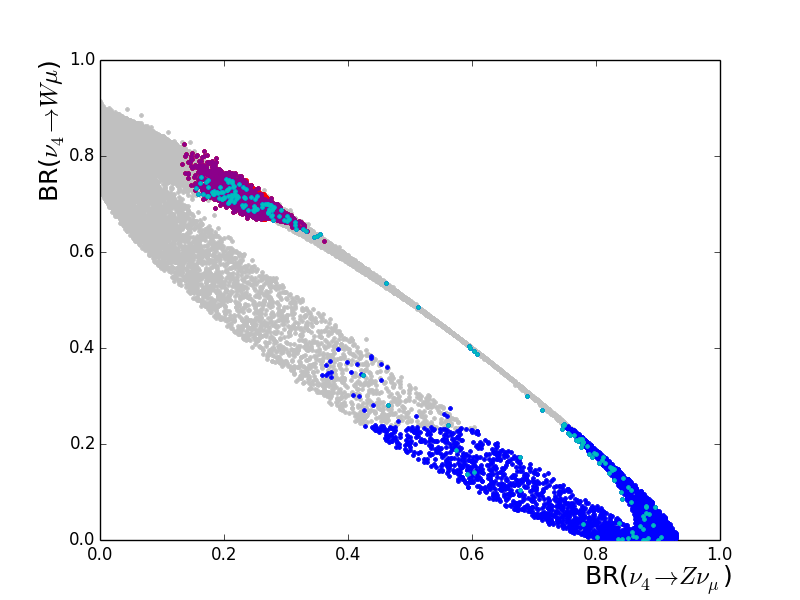}
\includegraphics[width=.49\linewidth]{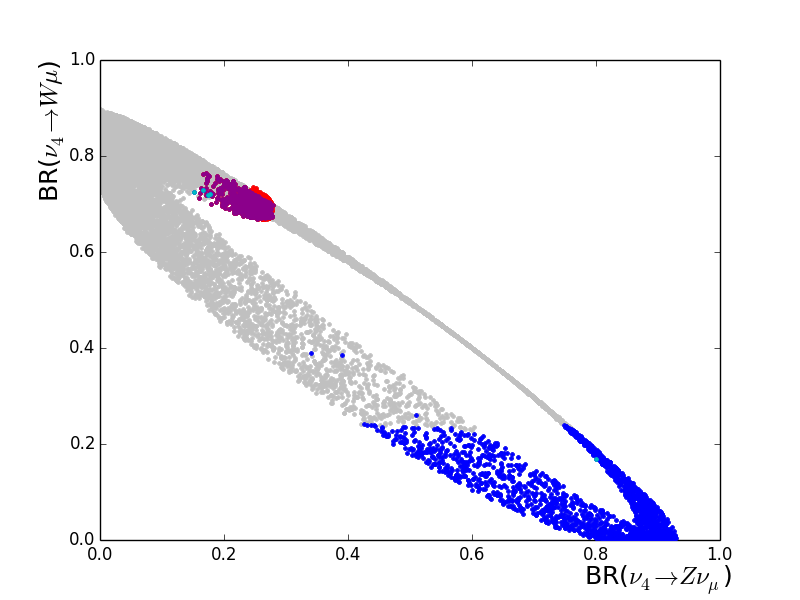}
\includegraphics[width=.49\linewidth]{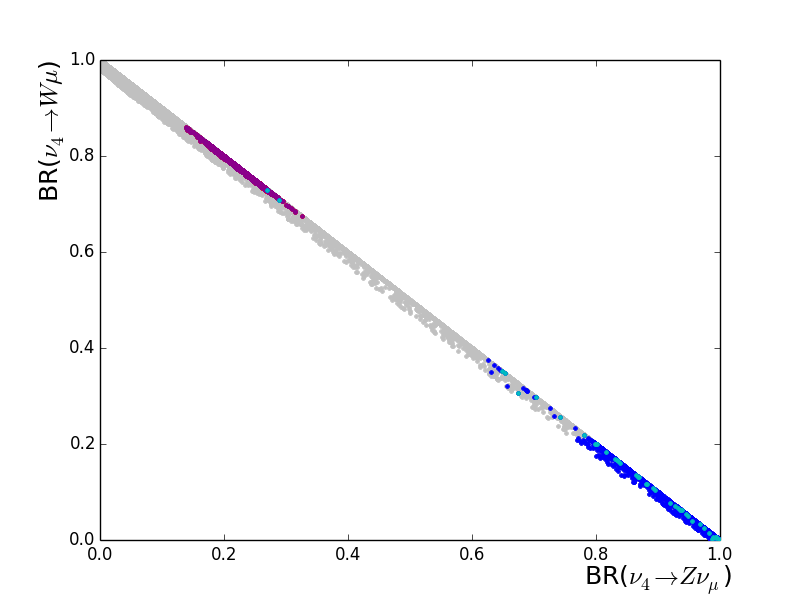}
\includegraphics[width=.49\linewidth]{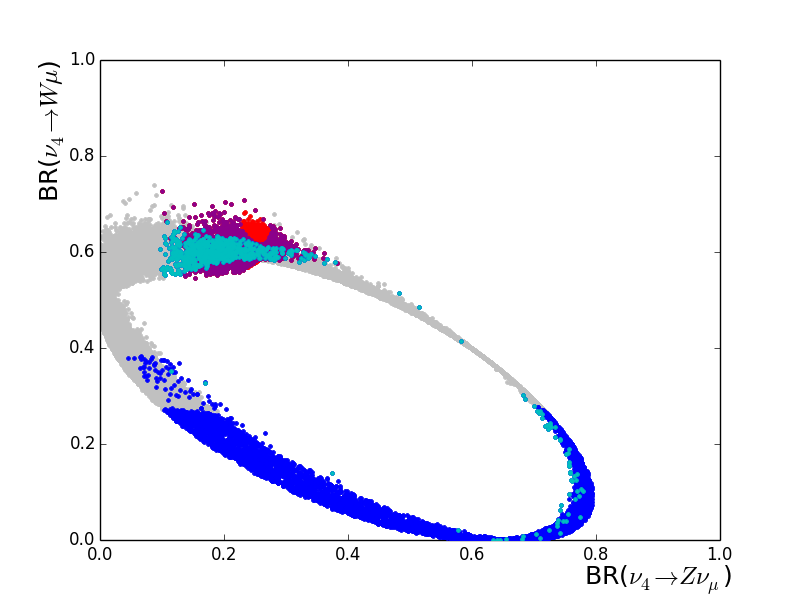}
\caption{Projection onto the $[{\rm BR}(\nu_4 \to Z \nu_\mu),{\rm BR}(\nu_4 \to W \mu )]$ plane of the parameter scan described in the main text for $(m_H,m_{\nu_4}) = (140\; {\rm GeV}, 135\; {\rm GeV}), (250\; {\rm GeV}, 135\; {\rm GeV}), (155\; {\rm GeV}, 125\; {\rm GeV})$ and $(155\; {\rm GeV}, 150\; {\rm GeV})$. See the caption in figure~\ref{fig:155135} for further details. \label{fig:brshelp}}
\end{center}
\end{figure}

In this section we explore the impact that this model has on $pp\to (WW, H\to WW) \to
\ell\nu\ell^\prime \nu^\prime$ measurements. We show detailed results for the two representative points $(m_H,m_{\nu_4}) = (155\; {\rm GeV},135\; {\rm GeV})$ and $(250\; {\rm GeV},230\; {\rm GeV})$ in figures.~\ref{fig:155135} and \ref{fig:250230}. In figures.~\ref{fig:mHmNemu}--\ref{fig:brshelp} we show how these results vary for different values of $m_H$ and $m_{\nu_4}$. 

In figure~\ref{fig:155135} we present the results of the scan described in section~\ref{sec:constraints} for the reference point $(m_H,m_{\nu_4}) = (155\; {\rm GeV},135\; {\rm GeV})$ discussed in ref.~\cite{Dermisek:2015vra}. The blue, cyan, magenta and red points have $\nu_4$ with singlet fraction ($| (V_L^\dagger)_{45}|^2 / 2 + |(V_R^\dagger)_{45}|^2 /2$) in the ranges [0,5]\%, [5,50]\%, [50,95]\%, and [95,100]\%, respectively (note that in some of these plots blue/cyan/magenta colors are not easily distinguishable). In the two upper plots the black contours are the values of the effective $pp\to WW$ cross section as defined in \cite{Dermisek:2015vra} for the $e\mu \nu_e \nu_\mu$ ($[\sigma_{\rm NP}^{WW}]_{e\mu}$) and $\mu\mu \nu_\mu \nu_\mu$ ($[\sigma_{\rm NP}^{WW}]_{\mu\mu}$) final states, respectively. In parenthesis we show the corresponding effective $pp\to H \to WW$ cross sections ($[\sigma_{\rm NP}^{H\to WW}]_{e\mu,\mu\mu}$). These effective cross sections 
\footnote{An extended discussion of the effective cross sections is presented in section 2 of ref.~\cite{Dermisek:2015vra}.}
are explicitly defined as:
\begin{align}
\sigma_{\rm NP} = \frac{\sigma (pp\to H \to W\ell\nu \to \ell\nu\ell^\prime\nu^\prime) \; A_{\rm NP}}{\eta \; {\rm BR} (W\to \ell\nu)^2\; A_{\rm SM}}~,
\label{effectiveXS}
\end{align}
where $\eta = 2 (1) $ for $e\mu$ ($\mu\mu$) final states and the NP and SM acceptances $A_{\rm NP}$ and $A_{\rm SM}$ are calculated using the experimental $WW$ and $H\to WW$ cuts (for the latter we follow ref.~\cite{Dermisek:2013cxa} and consider the six Higgs mass hypotheses discussed in ref.~\cite{cmshww} and show the most constraining effective cross section). Note that points displayed in the two upper panels are identical and that the only difference lies in the $\sigma_{\rm NP}^{WW}$ contours that depend crucially on the very different acceptances for $e\mu$ and $\mu\mu$ final states as well as the factor $\eta$. Note that eq.~(\ref{effectiveXS}) implies 
\begin{align}
\sigma_{\rm NP}^{H\to WW} & = \frac{A_{\rm NP}^{\cal H}}{A_{\rm SM}^{\cal H}} \frac{A_{\rm SM}^{WW}}{A_{\rm NP}^{WW}} \sigma_{\rm NP}^{WW} \; .
\end{align}
The product of acceptances in this equation is the crucial parameter that controls the size of contributions to $pp\to WW$ that are allowed by $H\to WW$ searches. For most (but not all) masses that we consider, this ratio is of order 10\% (typically $A_{\rm NP}^{\cal H}/A_{\rm SM}^{\cal H} \sim O(0.1)$ and $A_{\rm SM}^{WW}/A_{\rm NP}^{WW}\sim O(1)$). The smallness of $A_{\rm NP}^{\cal H}/A_{\rm SM}^{\cal H}$ is the reason for which we can find large $\sigma_{\rm NP}^{WW}$ cross sections while simultaneously surviving $H\to WW$ bounds. When the difference $m_H- m_{\nu_4}$ is large and $m_{\nu_4}$ is small, $E_T^{\rm miss}$ and $m_T$ increase while $m_{\ell\ell}$ decreases implying a larger $A_{\rm NP}^{\cal H}/A_{\rm SM}^{\cal H}$ ratio. For instance, for $m_H = 250\; {\rm GeV}$ and $m_{\nu_4} = 135\; {\rm GeV}$ we find that this ratio can be as large as 2. 

The yellow shaded area is excluded by $H\to WW$ searches. The upper bound on the effective $\sigma_{\rm NP}^{H \to WW}$ cross section is independent of $m_{\nu_4}$ and is given by
\dis{
\sigma_{\rm NP}^{H \to WW} < \min_{\mathcal H} \left[ \frac{\beta_{95}^{\mathcal H}}{A_{\rm SM}^{\mathcal H}} \right] \cdot \frac{1}{\eta \; {\rm BR} (W\to \ell\nu)^2}~,
}
where $\beta_{95}^{\mathcal H}$ is defined in appendix A of ref.~\cite{Dermisek:2015vra}.  The measurement of the $pp\to WW$ cross section is very sensitive to NNLO QCD corrections which have not been fully implemented in the experimental analysis yet. Following, for instance, the discussion around eq.~(1.7) of ref.~\cite{Dermisek:2015vra}, the deviation of the $pp\to WW$ cross section with respect to the SM expectation found by ATLAS~\cite{atlasww} and CMS~\cite{CMS:2015uda} are:
\begin{align}
\begin{cases}
[\sigma_{\rm NP}^{WW}]^{\rm ATLAS}_{e\mu} = \left(12.7^{+6.2}_{-5.8}\right)\; {\rm pb} \cr
[\sigma_{\rm NP}^{WW}]^{\rm ATLAS}_{\mu\mu} = \left(9.9^{+8.0}_{-7.3}\right)\; {\rm pb} \cr
\end{cases}
\;\; {\rm and} \;\;\;\;
\begin{cases}
[\sigma_{\rm NP}^{WW}]^{\rm CMS}_{e\mu} = (-0.1\pm 5.3)\; {\rm pb} \cr
[\sigma_{\rm NP}^{WW}]^{\rm CMS}_{\mu\mu} = (4.5\pm 7.0)\; {\rm pb} \cr
\end{cases} \; .
\label{eq:wwbounds}
\end{align}
Since these two results adopt different theoretical setups, we refrain from combining them into a weighted average. For this reason do not use $pp\to WW$ data to constrain our model and simply quote the allowed values.


\begin{figure}
\begin{center}
\includegraphics[width=.49\linewidth]{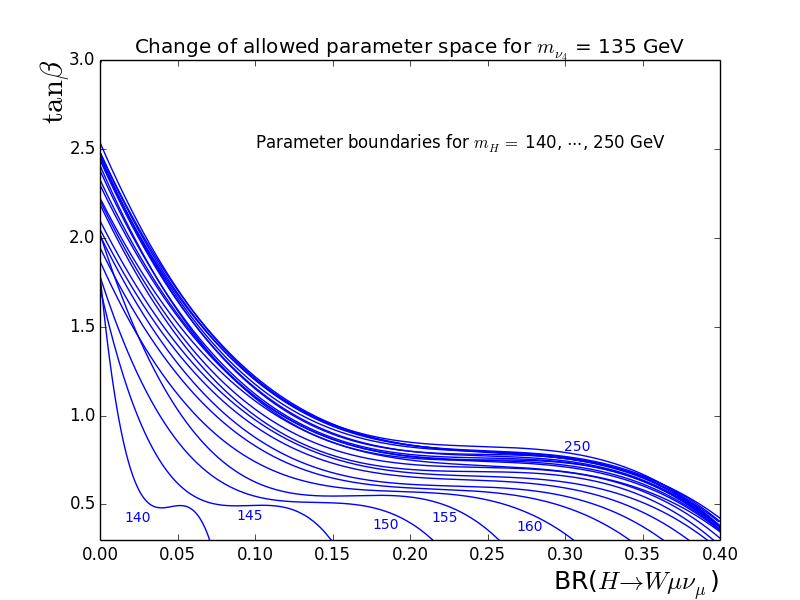}
\includegraphics[width=.49\linewidth]{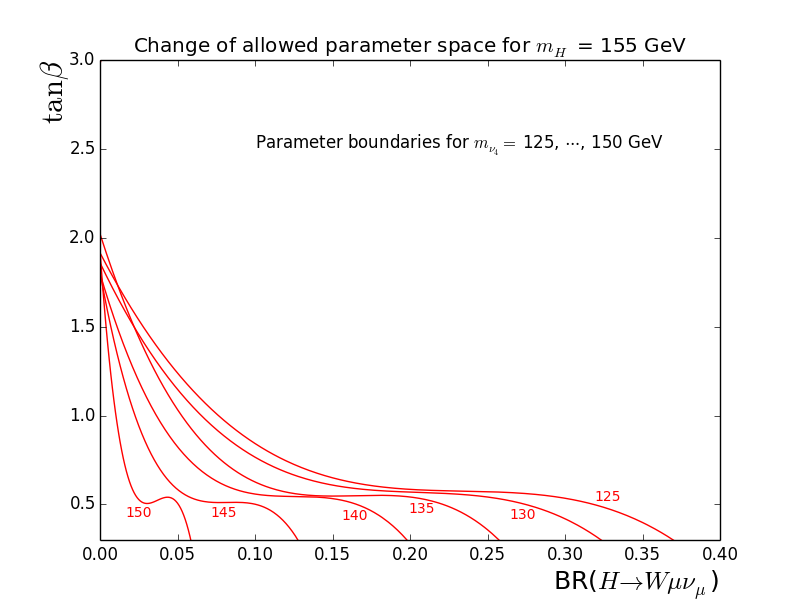}
\caption{In the left plot we show how the envelope of the points changes for fixed $m_{\nu_4} = 135 \; {\rm GeV}$ and $m_H \in [140,250]\; {\rm GeV}$. In the right plot we take $m_H = 155\; {\rm GeV}$ and $m_{\nu_4} \in [125, 150]\; {\rm GeV}$.  \label{fig:contours}}
\end{center}
\end{figure}

A prominent feature of figure~\ref{fig:155135} is that for a doublet-like $\nu_4$ the product of branching ratio ${\rm BR} (H\to W\mu\nu_\mu) = {\rm BR}(H \to \nu_4 \nu_\mu) \times {\rm BR} (\nu_4 \to W \mu)$ is constrained to be very small. This is mainly due to bounds from the multilepton plus $E_T^{\rm miss}$ searches in the Drell-Yan pair production process $pp \to \nu_4 \bar \nu_4 \to W^+ W^- \mu^+ \mu^-$. This can be understood by looking at the middle-left panel of figure~\ref{fig:155135} where we consider the $R_{\nu_4\nu_4} - {\rm BR} (\nu_4\to W\mu)$ plane. The quantity $R_{\nu_4\nu_4}$ is defined in eq.~(\ref{eq:rnn}). Here the light colored points are obtained without imposing any of the constraints discussed in section~\ref{sec:constraints} and the darker colored points are those that survive after imposing the muon lifetime bound.  Additional constraints from oblique corrections are very strong (especially from the $S$ parameter) but in the $[{\rm BR} (H\to W\mu\nu_\mu),\tan\beta]$ plane they do not modify significantly the overall allowed region.

Bounds from multilepton searches exclude the region above the black contour separating the surviving points in two disconnected regions at low $R_{\nu_4\nu_4}$ with ${\rm BR} (\nu_4\to W\mu) \sim 70\%$ and large $R_{\nu_4\nu_4} \sim 1$ with low ${\rm BR} (\nu_4\to W\mu)$.\footnote{Note that these arguments rely strongly on the particular choice of masses ($(m_H,m_{\nu_4}) = (155\; {\rm GeV},135\; {\rm GeV})$ in this case); a completely different situation characterizes the configuration presented in Fig~\ref{fig:250230} and discussed later on.}

At small $R_{\nu_4\nu_4}$ the $\nu_4$ is mostly singlet, the second term in eq.~(\ref{Zcoupling}) is suppressed by a factor $(V_L^\dagger)_{44}$ with respect to the first and the $\nu_4 - Z - \nu_\mu$ coupling is controlled by the single quantity $(V_L^\dagger)_{42}$ (we remind the reader that $(V_L)_{22}$ is very close to 1). Under the assumption of no mixing in the charged sector, the matrix $U$ is the identity and the $\nu_4 - W - \mu$ coupling in eq.~(\ref{Wcoupling}) is also controlled by the parameter $(V_L^\dagger)_{42}$. As a consequence the ratio of these two couplings is the same as in the SM (i.e. independent of flavor mixing parameters), implying an almost constant $\nu_4 \to W\mu$ branching ratio ($\sim 70\%$). At large $R_{\nu_4\nu_4}$ the $\nu_4$ is mostly doublet, both terms in the $\nu_4 - Z - \nu_\mu$ coupling are of similar size, and the $\nu_4 \to W\mu$ branching ratio can acquire any value depending on the choice of input parameters. On top of this one should note that the $\nu_4 \to h \nu_\mu$ channel is phase space suppressed for the case $m_{\nu_4} = 135\; {\rm GeV}$. These considerations are also illustrated in the middle-right plot of figure~\ref{fig:155135} where we show the points in the ${\rm BR}(\nu_4\to Z \nu_\mu) - {\rm BR}(\nu_4 \to W\mu )$ plane. Here the gray points are excluded by multilepton searches and, to a lesser extent, oblique corrections.

The surviving region at large $R_{\nu_4\nu_4}$ is also characterized by a very small $H-\nu_4-\nu_\mu$ coupling as we can see in the lower-left panel of figure~\ref{fig:155135}. In fact, an almost completely doublet $\nu_4$  requires very small couplings $\kappa$ and $\bar \kappa$, implying a strong suppression of the Yukawa coupling $\lambda^H_{\nu_4\nu_\mu}$ given, for doublet $\nu_4$, in eq.~(\ref{eq:lam24doublet}). This can be seen in the lower-right panel of figure~\ref{fig:155135} where we show the values of the Yukawa couplings $\lambda_{\nu_\mu \nu_4}^H$ and $\lambda_{\nu_\mu \nu_4}^h$ for the points that survive all constraints. Therefore BR($H \to \nu_4 \nu_\mu$), and hence BR($H \to W \mu \nu_\mu$), are very small for doublet-like $\nu_4$. If $\nu_4$ is singlet-like, the SM-like Higgs Yukawa coupling is given, in the limit of small mixing, by $\lambda_{\nu_\mu \nu_4}^h \sim \kappa_N \sin\beta$, see eq.~(\ref{eq:lam24singlet}). In this case, at fixed $M_N$, the muon lifetime limit (\ref{eq:kappa_N_limit}) translates into a direct constraint on the Yukawa coupling $\lambda_{\nu_\mu \nu_4}^h$.

In figure~\ref{fig:250230} we present the $(m_H,m_{\nu_4}) = (250\; {\rm GeV},230\; {\rm GeV})$ case. Now, the large $\nu_4$ mass implies that the decay mode $\nu_4 \to h \nu_\mu$ is no more phase space suppressed and can be dominant in large part of the parameter space as we can seen directly in the middle-right plot in figure~\ref{fig:250230} and indirectly in the middle-left plot where BR$(\nu_4 \to W \mu)$ can only be as large as 60\%. On top of this, the constraint from multilepton + $E_T^{\rm miss}$ searches is weaker (this happens generally for $m_{\nu_4} > 150$ GeV as we can see in Table~\ref{table:dynn}),  implying that there is a large region of allowed parameter space in which the $\nu_4$ is mostly doublet as can be seen in the two top plots in figure~\ref{fig:250230}. Even though there are many points for which the $\nu_4$ doublet fraction is large, the corresponding values for ${\rm BR}(H\to W\mu\nu)$ are much smaller than for typical singlet points.  This is because the $H -\nu_4- \nu_\mu$ coupling for a doublet-like $\nu_4$ is suppressed compared to the singlet-like $\nu_4$ by $\bar \kappa v_u / M_N$, see eqs.~(\ref{eq:lam24doublet}) and~(\ref{eq:lam24singlet}). From the bottom-right plot in figure~\ref{fig:250230} we see that the actual bounds on $\lambda_{\nu_\mu \nu_4}^h$ and $\lambda_{\nu_\mu \nu_4}^H$ are 0.05 and 0.17, respectively. Given that the ratio of these couplings is equal to $\tan\beta$, the second bound is effectively set by the perturbativity request $\tan\beta \gtrsim 0.3$.

In figures~\ref{fig:mHmNemu} (for the $e\mu$ final state) and \ref{fig:mHmNmumu} (for the $\mu\mu$ final state) we present the result of similar scans for $(m_H,m_{\nu_4}) = (140\; {\rm GeV}, 135\; {\rm GeV}), (250\; {\rm GeV}, 135\; {\rm GeV}), (155\; {\rm GeV}, 125\; {\rm GeV})$ and $(155\; {\rm GeV}, 150\; {\rm GeV})$. The interpretation of these plots is similar to that of figure~\ref{fig:155135}. The main difference between these plots is the maximum value allowed for ${\rm BR} (H\to W\mu\nu_\mu)$. In figure~\ref{fig:brshelp} we show the ${\rm BR}(\nu_4\to Z \nu_\mu) - {\rm BR}(\nu_4 \to W\mu )$ plane for each set of masses.

In figure~\ref{fig:contours} we show the envelopes of the allowed parameter space for a wide range of masses; in the left plot we take $m_{\nu_4} = 135 \; {\rm GeV}$ and $m_H \in [140,250]\; {\rm GeV}$ and in the right plot we have $m_H = 155\; {\rm GeV}$ and $m_{\nu_4} \in [125, 150]\; {\rm GeV}$. This effect is due to change in the phase space available for $H\to \nu_4 \nu_\mu \to W \mu \nu_\mu$ as the masses vary.

 Assuming that the acceptances ratios $A_{\rm NP}^{WW} / A_{\rm SM}^{WW}$ and $A_{\rm NP}^{\mathcal H} / A_{\rm SM}^{\mathcal H}$ remain constant when increasing the center of mass energy from 8 to 13 TeV, the $\sigma_{\rm NP}^{WW}$ contours in figures.~\ref{fig:155135}-\ref{fig:mHmNmumu} will simply scale with $pp\to H$ production cross section. For instance, for $m_H= 250\; {\rm GeV}$ the rescaling factor is about 2.776~\cite{Heinemeyer:2013tqa}. 

Finally let us comment on the reach of the next LHC run at 13 TeV with a luminosity ${\cal L} = 100 \; {\rm fb}^{-1}$. Taking into account that $\sigma(pp\to WW)_{13{\rm TeV}}^{\rm th}/\sigma(pp\to WW)_{8{\rm TeV}}^{\rm th} \simeq 2$~\cite{Gehrmann:2014fva} and that the uncertainty on $\sigma_{\rm NP}^{WW}$ is $\delta_{\rm exp}^{8 {\rm TeV}}  \simeq 5 \; {\rm pb}$ (see eq.~(\ref{eq:wwbounds})), we estimate $\delta_{\rm exp}^{13{\rm TeV}} \sim 3 \; {\rm pb}$. Moreover, our new physics contributions to $\sigma_{\rm NP}^{WW}$ scale with the $pp\to H$ cross section and increase by a factor $\sim 2.5$~\cite{Heinemeyer:2013tqa} at 13 TeV. Taking these considerations into account, direct inspection of figures.~\ref{fig:155135}-\ref{fig:mHmNmumu} shows that most of the presently allowed parameter space will be tested. For instance, with respect to the top-left panel of figure~\ref{fig:155135}, LHC8 with 20 ${\rm fb}^{-1}$ is sensitive to points below the 5 pb contour while LHC13 with 100 ${\rm fb}^{-1}$  will be sensitive to points roughly below the 1.2 pb one (that will correspond to $[\sigma_{\rm NP}^{WW}]_{13{\rm TeV}} \simeq 3 \; {\rm pb}$).

\section{Contributions from $H \to e_4 \mu$}
\label{sec:e4}
\begin{figure}
\begin{center}
\includegraphics[width=.49\linewidth]{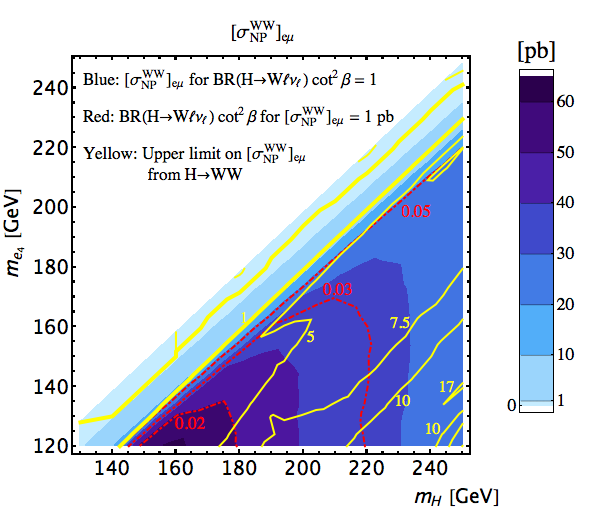}
\includegraphics[width=.49\linewidth]{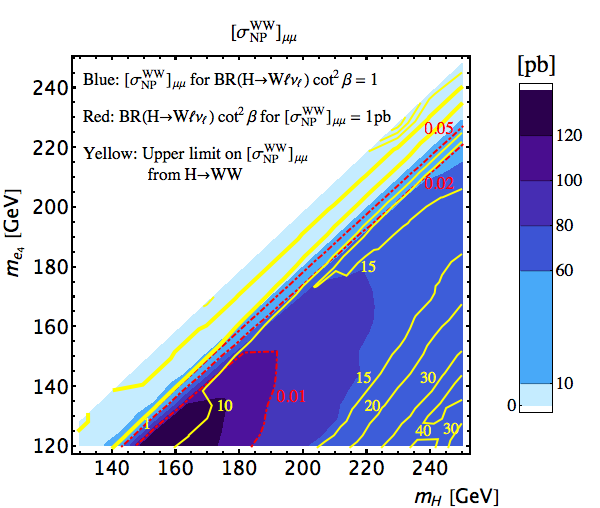}
\caption{The contours of $\sigma_{\rm NP}^{WW}$ for the process $H \to e_4 \mu \to W \mu \nu_\mu$ with BR($H \to W \mu \nu_\mu$) $\equiv$ BR($H \to e_4 \mu$) $\times$ BR($e_4 \to W \nu_\mu$) = 1 (this definition applies only here) for the $e\mu$ final state (left) and $\mu \mu$ final state (right). 
}
\label{fig:He4mu}
\end{center}
\end{figure}

In this section we discuss contributions to $pp\to \ell\ell^\prime \nu_\ell\nu_{\ell^\prime}$ stemming from heavy Higgs production and decay into a charged vectorlike lepton and a muon:
\begin{align}
pp \to H \to e_4 \mu \to W \nu_\mu \mu \to \mu \ell \nu_\mu \nu_\ell \; .
\end{align}
 We begin our analysis with a model independent study of this channel along the lines of the analysis presented in ref.~\cite{Dermisek:2015vra}. Our main results are summarized for the $e\mu$ and $\mu\mu$ modes in the two panels of figure~\ref{fig:He4mu}. These figures are very similar to figure 1 of ref.~\cite{Dermisek:2015vra}. The blue contours are the values of the effective $WW$ cross section $\sigma_{\rm NP}^{WW}$ that we obtain for ${\rm BR} (H \to W \mu \nu_\mu) \cot^2 \beta = 1$ (in this section only we define ${\rm BR}(H \to W \mu \nu_\mu) \equiv {\rm BR}(H \to e_4 \mu) \times {\rm BR}(e_4 \to W \nu_\mu)$). The yellow contours are the upper bounds on $\sigma_{\rm NP}^{WW}$ (in pb) implied by the $H\to WW$ limits and are controlled by the dependence of our signal acceptances (for the $WW$ and $H\to WW$ analyses) on the $H$ and $e_4$ masses. The red dashed contours are labelled with the value of ${\rm BR} (H \to W \mu \nu_\mu) \cot^2 \beta$ that leads to $\sigma_{\rm NP}^{WW} = 1 \; {\rm pb}$. 

Focusing on the $e\mu$ case (for which there is a larger statistics), we see that in the bulk of the parameter space we consider the maximum allowed $WW$ effective cross sections are smaller than 10 pb and well within the allowed $2 \sigma$ experimental ranges (see eq.~(\ref{eq:wwbounds})). This is in contrast to what happens for $H\to\nu_4 \nu_\mu$ as one can see from figure~1 of ref.~\cite{Dermisek:2015vra} where $H\to WW$ constraints allow for very large effective $WW$ cross sections in most of the parameter space. 

This feature is due to the different behavior of the ratio of acceptances $A_{\rm NP}^{WW}/A_{\rm NP}^{\cal H}$ for the $H\to \nu_4\nu_\mu$ and $H\to e_4 \mu$ channels. This ratio controls the upper limit on the effective $WW$ cross section (as we explain in appendix A of ref.~\cite{Dermisek:2015vra}, the larger the ratio, the larger the allowed cross section). Both channels have similar $A_{\rm NP}^{WW}/A_{\rm NP}^{\cal H}$ ratio at moderately large $m_H$ and small $m_{\nu_4,e_4}$; this implies that the 10 pb yellow contours for the $\nu_4$ and $e_4$ cases are close to each other. As we explain below, when moving to smaller $m_H$ and larger $m_{\nu_4,e_4}$ the $\nu_4$ ratio increases while the $e_4$ one decreases. Because of this behavior, in the bulk of the parameter space in which we are interested (smaller $m_H$ and larger $m_{\nu_4,e_4}$), we find large allowed $\sigma_{\rm NP}^{WW}$ values for the $H\to \nu_4\nu_\mu$ channel but not for the $H\to e_4\nu_\mu$ one.

The behavior of the acceptances ratio is essentially controlled by the difference $m_H - m_{\nu_4,e_4}$. This difference determines the transverse mass $m_T$ in the $\nu_4$ case and the dilepton invariant mass $m_{\ell\ell}$ in the $e_4$ one. The $H\to WW$ acceptance decreases for channels with lower $m_T$ and increases for channels with lower $m_{\ell\ell}$ (because the CMS Higgs cuts include a range for $m_T$ and an upper bound on $m_{\ell\ell}$). The $WW$ acceptance, on the other hand, is controlled by a $m_{\ell\ell}> 10 (15) \; {\rm GeV}$ cut (for $e\mu$ and $\mu\mu$ final states): for the $e_4$ case it decreases at low $m_H - m_{e_4}$, while, for the $\nu_4$ one, the dilepton invariant mass is controlled by $m_{\nu_4}$ and tends to always pass the cut implying a mild dependence of the $WW$ acceptance on the choice of masses. In conclusion, small $m_H - m_{\nu_4,e_4}$ implies a small acceptances ratio for $e_4$ and a large one for $\nu_4$.

The discussion of the $\mu\mu$ mode is similar. The main differences are that the experimental $H\to WW$ cuts are much tighter in order to suppress Drell-Yan backgrounds and that the effective cross section is enhanced by a combinatorial factor of 2 with respect to the $e\mu$ case  (see eq.~(\ref{effectiveXS})). Note that in order to obtain similar effective cross sections for the $e\mu$ and $\mu\mu$ modes one needs to include a second vectorlike lepton family as discussed in section 4 of ref.~\cite{Dermisek:2015vra}.

Since the $WW$ effective cross sections that we find in the $H\to e_4\mu$ channel are typically smaller than 10 pb, we refrain from performing a detailed scan that includes mixing in the charged lepton sector.

\begin{figure}
\begin{center}
\includegraphics[width=.49\linewidth]{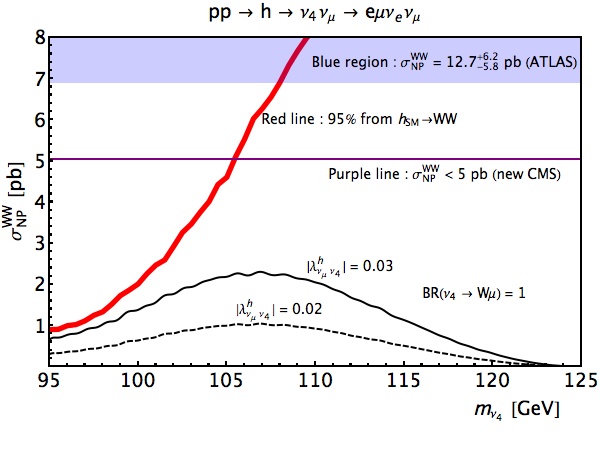}
\includegraphics[width=.49\linewidth]{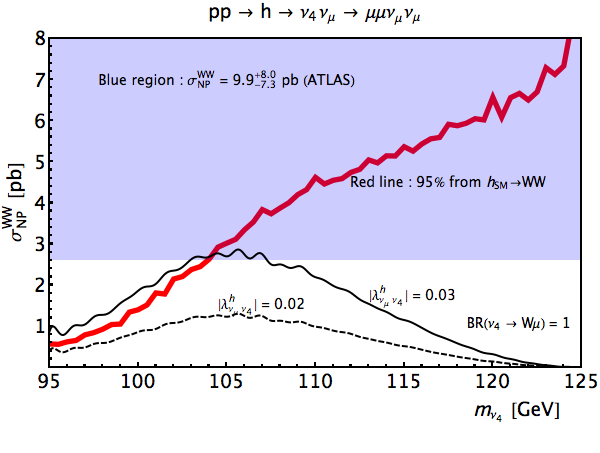}
\caption{Contributions of the 125 GeV Higgs decay $h \to \nu_4 \nu_\mu$ to the effective cross section $\sigma_{\rm NP}^{WW}$ as a function of $m_{\nu_4} \in [95, 125]$ GeV for the $e\mu$ and $\mu\mu$ modes. We set BR($\nu_4 \to W \mu$) = 1 and consider two representative values of the flavor violating Yukawa couplings $|\lambda_{\nu_\mu \nu_4}^h| = 0.02$ and 0.03. The thick solid red line in figure~\ref{fig:h125} is the 95\% C.L. upper bound from the SM $h \to WW$ ATLAS search~\cite{ATLAS:2014aga}. \label{fig:h125}}
\end{center}
\end{figure}
\begin{figure}
\begin{center}
\includegraphics[width=.49\linewidth]{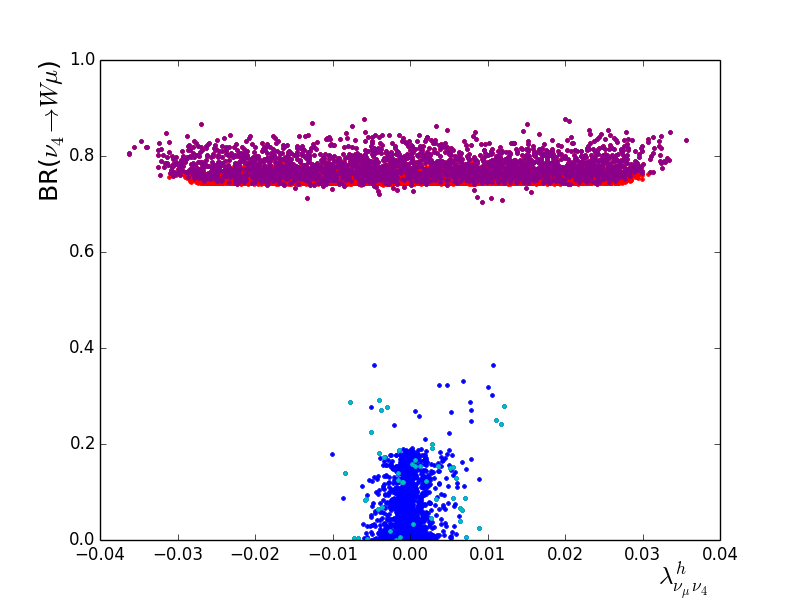}
\caption{Result of a parameter scan in the $\lambda_{\nu_\mu \nu_4}^h$ - BR($\nu_4 \to W \mu$) plane for $m_{\nu_4} = 120$ GeV. Similar results are obtained for other choices of $m_{\nu_4} < 125\; {\rm GeV}$. See the caption in figure~\ref{fig:155135} for further details on the scan.}
\label{fig:yukawamn120}
\end{center}
\end{figure}

\section{Contributions from SM-like Higgs boson }
\label{sec:SMh}
In this section we consider exotic decays of the SM-like Higgs into vectorlike leptons. 
We begin by considering the $h \to \nu_4 \nu_\mu \to W \mu \nu_\mu$ process. In figure~\ref{fig:h125} we show the effective cross sections $\sigma_{\rm NP}^{WW}$ as a function of $m_{\nu_4} \in [95, 125]$ GeV for the $e\mu$ and $\mu\mu$ modes. Here we set BR($\nu_4 \to W \mu$) = 1 and consider two representative values of the flavor violating Yukawa couplings $|\lambda_{\nu_\mu \nu_4}^h| = 0.02$ and 0.03.  These values are close to the largest possible as one can see from the parameter scan presented in figure~\ref{fig:yukawamn120} where we show the $|\lambda_{\nu_\mu \nu_4}^h|  - {\rm BR}(\nu_4 \to W \mu)$ plane for $m_{\nu_4} = 120\; {\rm GeV}$. The thick solid red line in figure~\ref{fig:h125} is the 95\% C.L. upper bound from the SM $h \to WW$ ATLAS search~\cite{ATLAS:2014aga}. 

We see that, for the $e\mu$ mode, Higgs searches are not constraining while in the $\mu\mu$ mode they require $m_{\nu_4} \gtrsim 105$ GeV. In both cases, the effective $WW$ cross section cannot exceed $2-3$ pb. These cross sections are far from the ranges allowed by ATLAS (blue shaded region) and the CMS upper bound (purple line) for the $e\mu$ final state but are close to the allowed ATLAS region for the $\mu\mu$ case (see eq.~(\ref{eq:wwbounds}) and the related discussion). 

We do not discuss in detail the $h \to e_4 \mu \to W \mu \nu_\mu$ process because we found that it does not lead to appreciably large effective cross sections (typically smaller than $1 {\rm pb}$) and it is severely constrained by $h \to WW$ searches.

\section{Contributions from Drell-Yan production of vectorlike leptons}
\label{sec:DY}
In this section we discuss contributions to the effective cross section $\sigma_{\rm NP}^{WW}$ that stem from the following vectorlike lepton Drell-Yan production processes ($\ell=e,\; \mu$):
\begin{align}
&p p \to (\gamma,Z) \to e_4^\pm e_4^\mp \to W^\pm W^\mp \nu_\mu \bar{\nu}_\mu \to 2\ell 4\nu~,
\label{eq:DYrelevant1}\\
&p p \to Z \to \nu_4 \nu_\mu \to W \mu \nu_\mu \to \ell \mu 2\nu~.
\label{eq:DYrelevant2}
\end{align}
Note that there are many more processes (involving up to four light leptons in the final state) that one can consider and the two modes we consider in eqs.~(\ref{eq:DYrelevant1}) and (\ref{eq:DYrelevant2}) are the two most promising ones.\footnote{If more than two light charged leptons are present, the third hardest lepton must have $p_T < 7 \; {\rm GeV}$ in order to avoid detection and this requirement suppresses the acceptance.}
\begin{figure}
\begin{center}
\includegraphics[width=.49\linewidth]{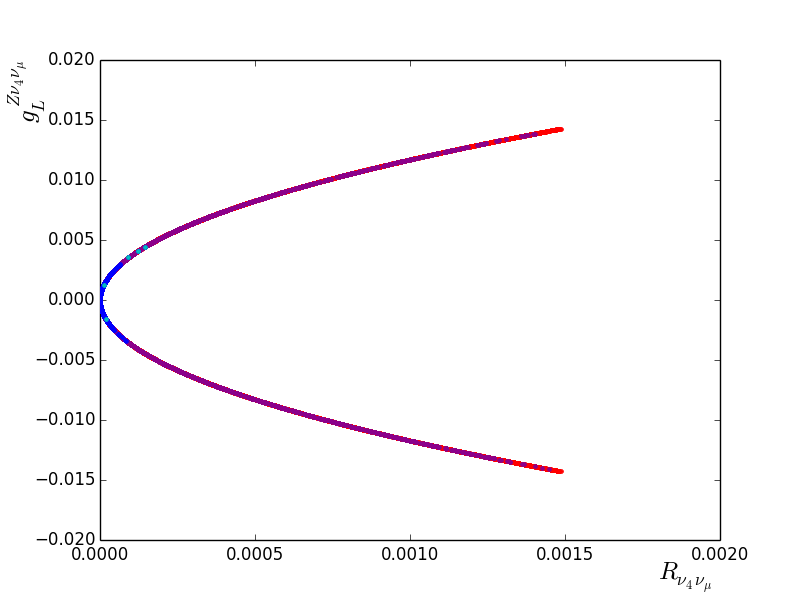}
\caption{Allowed values of $R_{\nu_4 \nu_\mu}$ and $g_L^{Z \nu_4 \nu_\mu}$ for $m_{\nu_4} = 110$ GeV. We see that $R_{\nu_4 \nu_\mu} \lesssim 1.5 \times 10^{-3}$ and $|g_L^{Z \nu_4 \nu_\mu}| \lesssim 0.02$. Similar bounds are found for different $\nu_4$ masses.}\label{fig:single2l}
\end{center}
\end{figure}

The $e_4 e_4$ pair production channel is flavor diagonal and the $Z-e_4-e_4$ coupling can be as large as the corresponding $Z-\ell-\ell$ SM one. On the other hand, in the channel (\ref{eq:DYrelevant2}) the production of a single $\nu_4$ is constrained by the values of the off-diagonal coupling $g_L^{Z \nu_4 \nu_\mu}$ allowed by EW precision data. To quantify this effect we define  
\begin{align}
R_{\nu_4 \nu_\mu} &\equiv \frac{(g_L^{Z \nu_4 \nu_\mu})^2}{g^2 / (4 \cos^2\theta_W)}~,
\end{align}
which shows how the production of $\nu_4 \nu_\mu$ through $Z$ boson is suppressed compared to the SM process $p p \to Z \to \nu_\mu \nu_\mu$. In Fig. \ref{fig:single2l} we see that $R_{\nu_4 \nu_\mu}$ can be at most $1.5 \times 10^{-3}$ for $m_{\nu_4} = 110$ GeV.

\begin{figure}
\begin{center}
\includegraphics[width=.49\linewidth]{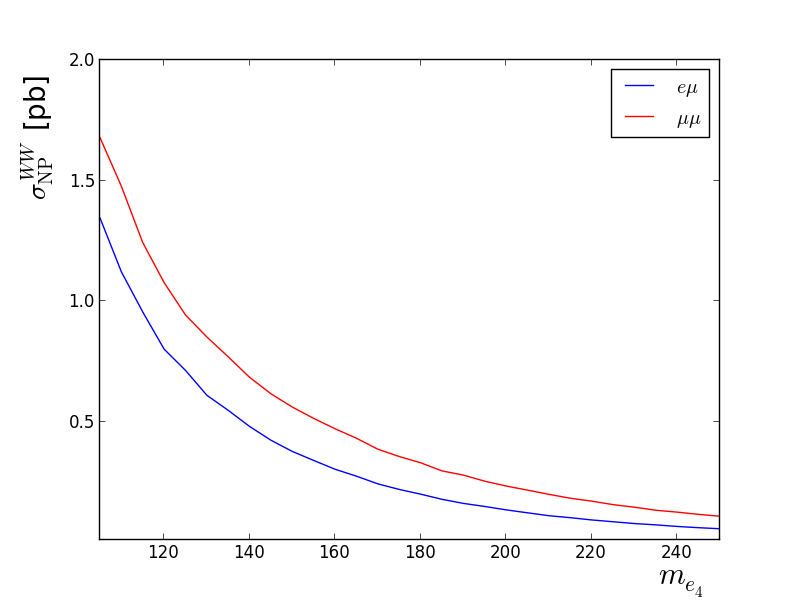}
\includegraphics[width=.49\linewidth]{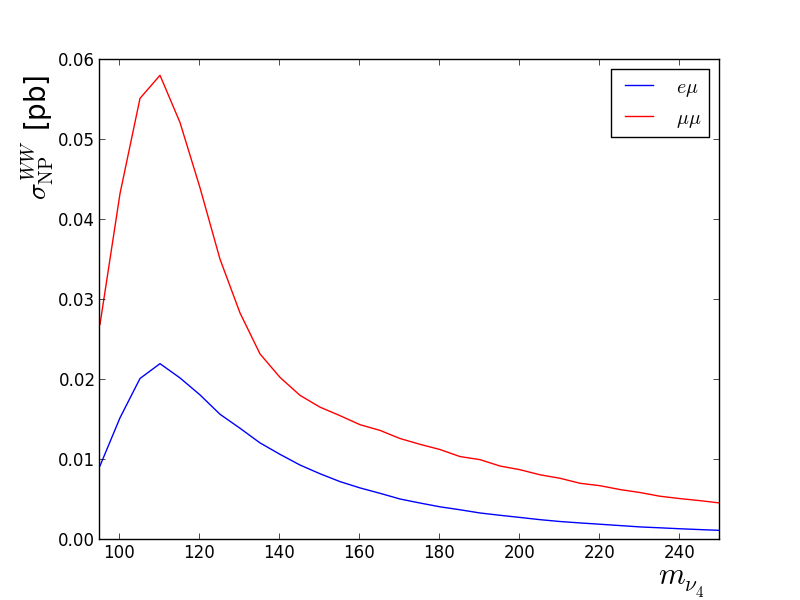}
\caption{The effective cross section $\sigma_{\rm NP}^{WW}$ [pb] for Drell-Yan processes. In the left panel we consider the channel $p p \to (\gamma,Z) \to e_4^\pm e_4^\mp \to W^\pm W^\mp \nu_\mu \bar{\nu}_\mu \to 2\ell 4\nu$ assuming SM-like strength of the $Z-e_4-e_4$ vertex, ${\rm BR}(e_4 \to W \nu_\mu) = 1$ and $m_{e_4} = 105 - 250$ GeV. In the right panel we show $p p \to Z \to \nu_4 \nu_\mu \to W \mu \nu_\mu \to \ell \mu 2\nu$ for $R_{\nu_4 \nu_\mu} \cdot {\rm BR}(\nu_4 \to W \mu) = 10^{-3}$ and $m_{\nu_4} = 95 - 250$ GeV.}
\label{fig:dy2l}
\end{center}
\end{figure}

In figure~\ref{fig:dy2l} we show the most optimistic values of effective cross sections for the processes in eqs.~(\ref{eq:DYrelevant1}) and (\ref{eq:DYrelevant2}). For the latter case we set $R_{\nu_4 \nu_\mu} \cdot {\rm BR}(\nu_4 \to W \mu) = 10^{-3}$. Consequently one can see that allowed values of $\sigma_{\rm NP}^{WW}$ for the $pp\to (\gamma,Z)\to e_4 e_4$ and $pp\to Z\to\nu_4 \nu_\mu$ channels are at most of order $1 \; {\rm pb}$ and $0.1 \; {\rm pb}$ respectively.

\section{Conclusions}
\label{sec:conclusions}

We studied decay modes of a heavy CP even Higgs boson, $H \to \nu_4 \nu_\mu$ and $H \to e_4 \mu$ followed by $\nu_4 \to W \mu$ and $e_4 \to W  \nu_\mu$, where $e_4$ and $\nu_4$ are the  lightest charged and neutral mass eigenstates originating from vectorlike pairs of SU(2) doublet and singlet new leptons.  We showed that, with Yukawa couplings as in two Higgs doublet model type-II, these decay modes, when kinematically open, can be large or even dominant. After imposing all the experimental constraints, the
$H \to \nu_4 \nu_\mu$  decay channel can have branching ratio of up to about 35\%.  

 As we discussed in sections~\ref{sec:constraints} and \ref{sec:results}, electroweak precision data impose very strong bounds on various gauge and Yukawa couplings: the new flavor violating gauge couplings $g_{L,R}^{W \nu_4 \mu}$ and $g_{L}^{Z \nu_4 \nu_\mu}$ have to be smaller than $\mathcal{O}(10^{-2})$, the couplings of SM gauge bosons to the second family of leptons, $g_L^{W \nu_\mu \mu}$ and $g_L^{Z \nu_\mu \nu_\mu}$, can deviate from their SM values by less than $\sim 0.1$\%, and the flavor violating Yukawa couplings $\lambda_{\nu_\mu \nu_4}^h$ and $\lambda_{\nu_\mu \nu_4}^H$ are constrained to be smaller than $\sim 0.05$ and $\sim 0.17$, respectively.

Focusing on $pp \to H \to \nu_4 \nu_\mu \to W \mu \nu_\mu$ we studied possible effects of this process on the measurements of $pp \to WW$ and $H \to WW$. Contributions from this process to $2\ell 2\nu$ final states can be very large since only one $W$ has to decay to leptons unlike in the case of $pp \to WW$ and $H \to WW$. We present predictions of the model in terms of effective cross sections for $pp \to WW$ and $H \to WW$ in $\mu e 2\nu$ $2\mu 2\nu$ final states from the region of the parameter space that satisfies all available constraints including precision electroweak observables and constraints from pair production of vectorlike leptons. Parts of the parameter space are already excluded by these measurements and thus possible contributions to these processes can be as large as current experimental limits. Large contributions, close to current limits, favor small $\tan \beta $ region of the parameter space.

In addition, we studied correlation of  the contributions to $pp \to WW$ and $H \to WW$.  We showed that, as a result of adopted cuts in experimental analyses,   the contribution to  $pp \to WW$  can be more than an order of magnitude larger than the contribution to $H \to WW$. Thus more precise measurement of $pp \to WW$ in future will significantly constrain the parameter space of the model. 
 
Furthermore, we also considered  possible contributions to $pp\to WW$ from $H\to e_4 \mu \to W \mu \nu_\mu$, from similar processes involving  SM-like Higgs boson and from pair production of vectorlike leptons. These however lead to much smaller contribution to the effective cross section for $pp \to WW$ while satisfying limits from $H \to WW$ and $h \to WW$ in first two cases. In the case of pair production of vectorlike leptons, the cross sections are very small, and the contribution to the effective $pp\to WW$ is at most of order 1pb.

 Finally, as we discussed at the end of section~\ref{sec:results}, the next LHC run at 13 TeV with $100 \; {\rm fb}^{-1}$ of integrated luminosity will be able to explore most of the parameter space currently allowed by electroweak precision data and $H\to WW$ constraints.

\acknowledgments 
RD thanks Hyung Do Kim and Seoul National University for kind hospitality during final stages of this project. 
The work of RD was supported in part by the U.S. Department of Energy under grant number {DE}-SC0010120, by the Munich Institute for Astro- and Particle Physics (MIAPP) of the DFG cluster of excellence ``Origin and Structure of the Universe" and by  the Ministry of Science, ICT and Planning (MSIP), South Korea, through the Brain Pool Program.

\end{document}